\documentclass[two column,showkeys]{revtex4}
\usepackage[T1]{fontenc}
\usepackage[latin1]{inputenc}
\usepackage{graphicx}
\usepackage{amsmath}
\usepackage{amssymb}
\usepackage{amsfonts}
\usepackage{color}
\usepackage{cancel}
\def\a{\alpha}
\def\b{\beta}
\def\g{\gamma}
\def\e{\varepsilon}
\def\d{\delta}

\def\l{\lambda}
\def\m{\mu}

\def\t{\tau}

\def\o{\omega}
\def\r{\rho}
\def\s{\sigma}
\def\E{\mathcal{E}}
\def\S{\Sigma}
\def\G{\Gamma}
\def\D{\Delta}

\def\ba{{\bf a}}
\def\bk{{\bf k}}
\def\bK{{\bf K}}
\def\bq{{\bf q}}
\def\br{{\bf r}}

\def\bn{{\bf n}}

\def\bde{{\boldsymbol{\delta}}}
\def\be{\begin{equation}}
\def\ee{\end{equation}}
\def\bea{\begin{eqnarray}}
\def\eea{\end{eqnarray}}
\def\llang{\langle\langle}
\def\rrang{\rangle\rangle}
\def\nn{\nonumber}
\def\lb{\label}

\begin{document}

\title{Impurity resonance effects in graphene {\it vs} impurity location, concentration and
sublattice occupation}

\author{Yuriy G. Pogorelov}%
	\email{ypogorel@fc.up.pt}
 	\affiliation{IFIMUP-IN,~Departamento~de~F\'{i}sica,~Universidade~do~Porto,~Porto,~Portugal,}

\author{Vadim M. Loktev}%
    \email{vloktev@bitp.kiev.ua}
 	\affiliation{N.~N.~Bogolyubov~Institute~of~Theoretical~Physics,~NAS~of~Ukraine,~Kyiv,~Ukraine, \\
 	\&\\
 	Igor~Sikorsky~Kyiv~Polytechnic~Institute,~Kyiv,~Ukraine,}
 	
\author{Denis Kochan}%
    \email{denis.kochan@ur.de}
 	\affiliation{Institute~for~Theoretical~Physics,~University~of~Regensburg,~Regensburg,~Germany}

\begin{abstract}
Unique electronic band structure of graphene with its semi-metallic features near the charge neutrality point is sensitive
to impurity effects. Using the Lifshitz and Anderson impurity models, we study in detail the disorder induced spectral
phenomena in the electronic band structure of graphene, namely, the formation of resonances, quasi-gaps, bound states,
impurity sub-bands, and their overall impact on the electronic band restructuring and the associated Mott-like metal-insulator
transitions. We perform systematic analytical and numerical study for realistic impurities, both substitutional and adsorbed,
focusing on those effects that stem from the impurity adatoms locations (top, bridge, and hollow positions), concentration,
host sublattice occupation, perturbation strengths, etc. Possible experimental and practical implications are discussed as
well.
\end{abstract}

\date{\today}
\keywords{graphene, impurity resonance, Lifshitz model, Anderson model, localization, Ioffe-Regel-Mott~criterium, group expansion}
\maketitle

%---------------------------------------------------------------
\section{\label{sec:intr}Introduction}
%---------------------------------------------------------------

Graphene is the first two-dimensional crystal possessing linear dispersion of low energy electronic states. Therefore they can be
described by an effective Dirac equation for 2D massless fermions. However, in experiments long-range Coulomb scattering
\cite{Adam2007,Swartz2013,Jia2015,Chandni2015} off
charged adatoms, as well, short-range scattering off the non-charged impurities can strongly affect graphene's
transport properties. A representative example of a short-range impurity is vacancy that is predicted to give rise
to zero energy resonance states in graphene \cite{Pereira2006,Pereira2008,Nanda2012}. Due to the small density of
states (DOS) at low energy, graphene is especially very sensitive to such induced resonant states
\cite{Stauber2007,Ferreira:PRB2011,Monteverde2010,Robinson2008,PhysRevB.99.035412}. Another source for these states are various substitutional
impurities \cite{Basko2008,Wehling2007,Pereira2008,Skrypnyk2006} or adsorbates in graphene. The latter have
been studied for specific adatoms by explicit tight-binding and density-functional theory calculations, see
for example \cite{Ihnatsenka:PRB2011,Wehling:PRB2010,Wehling:PRL2010,Wehling:PRB2007,Farjam2011,Gmitra:PRL2013,Zollner:Meth2016,Frank:PRB2017},
It was also realized by the basic symmetry analysis that the
adsorption position of an adatom plays an important role for the resonance scattering mechanism
\cite{Ruiz2016,Uchoa2014,Weeks2011,Duffy2016,PhysRevB.97.075417}. For example, it was established that the $s$-orbital of an adatom in hollow
position is effectively decoupled from the electronic states of graphene \cite{Ruiz2016} so that resonance scattering of
such an orbital is strongly suppressed. Generally, this sensitivity to impurity location can be related to the
specifics of graphene lattice that owns two sublattices, each of them with no local inversion symmetry, the same that defines the most
notable feature of pure graphene's spectrum, its Dirac points.

Our work aims to provide extended, self-contained and systematic study of spectral properties of graphene in the presence of
impurity disorder and the underlying onset of the Mott-like metal insulator transitions considering dependencies on impurity
concentration, their position type (top, bridge, hollow), sublattice occupation asymmetry and so on, and connect those with
some previous theoretical studies available in the literature. We consider two models; Lifshitz isotopic model \cite{Lifshitz},
and Anderson hybrid model \cite{Anderson1}. As will be shown, distribution of impurities position with respect to the host
sublattices can create an occupational asymmetry. The spectral properties of graphene (resonances, quasi-gaps, mobility edges,
impurity subgaps, etc.) are very sensitive, besides the total impurity concentration, also to such partial occupation asymmetries.

The paper is organized as follows, Section \ref{sec:model} presents a short introduction into the formulation of the tight-binding model and Green's functions formalism. Then in Sec. \ref{Lif} we consider the simpler Lifshitz isotopic model of
impurity perturbation and demonstrates certain specific effects appearing there even in the absence of impurity
resonances. Those resonances in their general form are further investigated in Sec. \ref{And} within the scope of Anderson's hybrid model, while 
Secs.~\ref{pos} and \ref{pos2} analyzes their particular realizations for different types of impurity
positions and their sublattice occupations. Finally, a discussion of the obtained results and their possible applications
are given in Sec. \ref{Disc}. Some more technical details of calculations, such as the restructured spectrum to higher order,
are provided in Appendices \ref{A} and \ref{B}.

\section{\label{sec:model}Model and Green functions}

We model the unperturbed graphene in terms of the tight-binding Hamiltonian:
\be
H_0 = t \sum_{\langle \bn_1,\bn_2\rangle} \left(b_{\bn_1}^\dagger b_{\bn_2}^{\phantom{\dagger}}
+ h.c.\right)\,,
\lb{H0}
\ee
where the carbon $2p_z$-atomic level is chosen as the energy reference. Hoppings, parameterized by the amplitude $t$,
connect nearest-neighbor graphene sites as symbolically indicated by $\langle \bn_1,\bn_2\rangle$. Here and below
$\bn_1$ stands for a site from sublattice-type 1 (A-sublattice), and $\bn_2$ for sublattice-type 2 (B-sublattice),
see Fig.~\ref{fig1}, in generic case we use symbol $\bn_j$. Here and in what follows we do not consider explicitly
the electron spin degrees of freedom assuming purely spin-diagonal hoppings so that the on-site energies and all
the observable quantities are understood per single spin projection.

The Hamiltonian $H_0$ is routinely diagonalized passing from the direct-space representation, through the local
atomic Fermi operators $b_{\bn_j}^{(\dagger)}$, to the corresponding Bloch band representation:
\begin{figure}
	\includegraphics[scale=0.8]{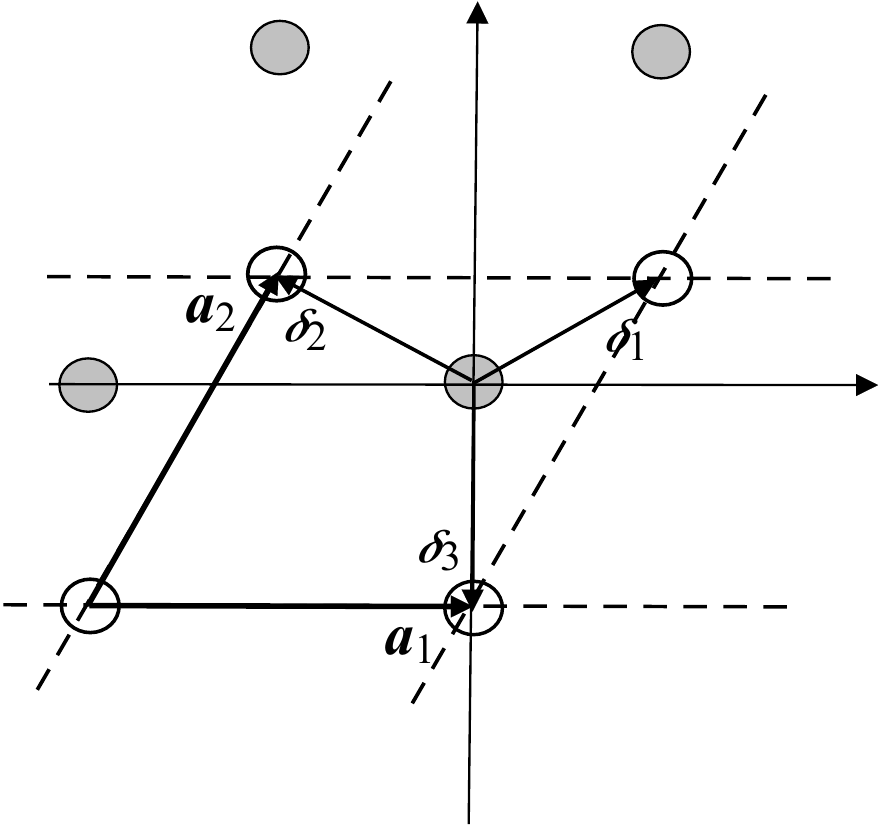}%Fig1
	\caption{Graphene lattice structure with $t$-hopping links along the nearest-neighbor vectors ${\boldsymbol
	\d}_{1,2,3}$ 	connecting type-1/sublattice-A (grey), and type-2/sublattice-B (white) carbon sites. Dashed lines
	mark the unit cells 	formed by the elementary translation vectors $\ba_{1,2}$.}
	\lb{fig1}
\end{figure}
\be
H_0 = \sum_{\bk} \e_\bk \left(\b_{+,\bk}^\dagger \b_{+,\bk}^{\phantom{\dagger}} - \b_{-,\bk}^\dagger \b_{-,\bk}^
{\phantom{\dagger}}\right).
\lb{Hb}
\ee
Here the eigenenergies $\e_\bk$, and the band operators $\b_{\pm,\bk}^{(\dagger)}$ are labeled by the wave-vector $\bk$
that belongs to the first Brillouin zone (BZ) spanned by the reciprocal basis vectors ${\bf b}_{1,2}$, i.e.~${\bf b}_j\cdot
\ba_{j'} = 2\pi \d_{j,j'}$, see Fig.~\ref{fig2}. Moreover, the sign subscript $\pm$ refers to the conduction and valence
bands, respectively. The corresponding energy dispersion laws, $\pm\e_\bk = \pm t|\g_\bk|$, follow from the hopping
factor:
\be
\g_\bk = \sum_{\boldsymbol \d} {\rm e}^{i\bk\cdot \boldsymbol \d} = 2\cos\frac{k_x}2{\rm e}^{i k_y/2\sqrt3} + {\rm e}
^{-i k_y/\sqrt3}\,,
\lb{GrHopFac}
\ee
(in what follows the quasi-momenta are measured in units of the inverse graphene lattice constant $a^{-1} = |\ba_{1,2}|^{-1}
= |\sqrt{3}\,\boldsymbol \d|^{-1}$).

The band, and the lattice (local atomic) operators are related via the Fourier transformation:
\be
b_{\bn_j} = \frac 1{\sqrt{2N}} \sum_\bk {\rm e}^{i\phi_{\bn_j,\bk}}\left(\b_{+,\bk} - (-1)^j\b_{-,\bk}\right),
\lb{ft}
\ee
where $N$ represents the number of unit cells, and the hopping phase reads:
\be
\phi_{\bn_j,\bk} = \bk\cdot\bn_j - \frac {(-1)^j }2 \arg \g_\bk\,.
\lb{Ang}
\ee

\begin{figure}
    \centering
	\includegraphics[scale=0.8]{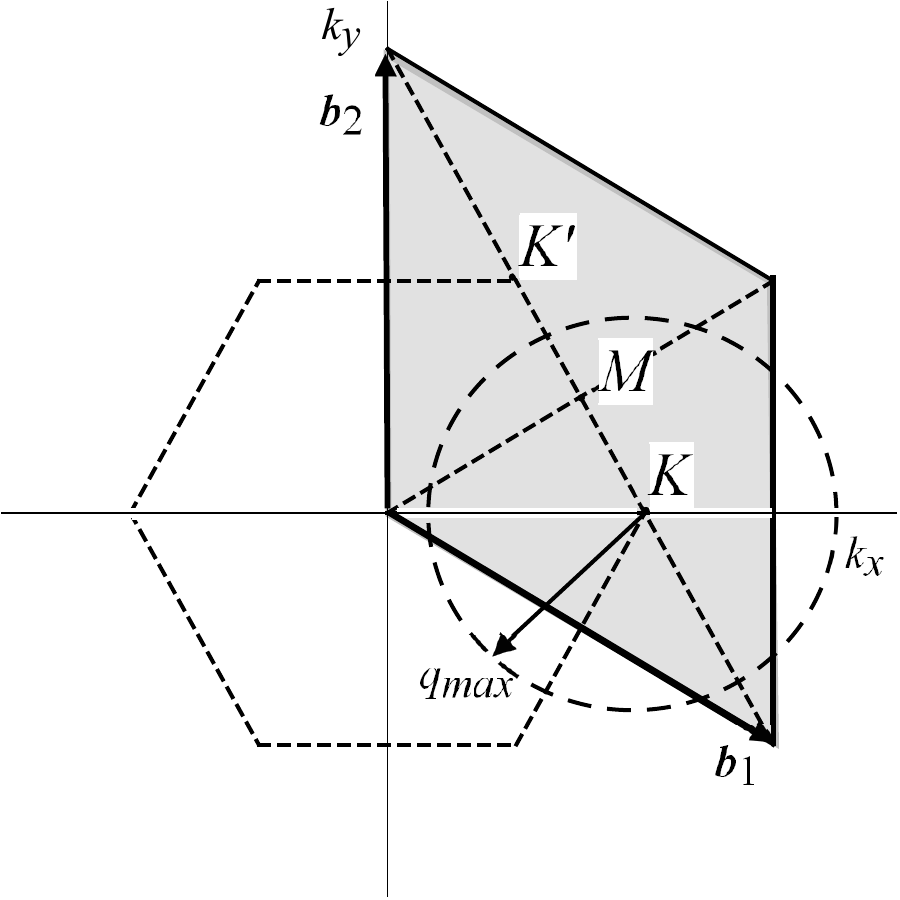}%Fig2
	\caption{Brillouin zone of graphene (shadowed rhombus) with the corresponding Dirac points $\mathbf K$ and
	$\mathbf K'$, and the associated reciprocal lattice vectors $\boldsymbol{b}_{1,2}$. An effective "half" of
	the Brillouin zone centered at $\bK$ point (dashed circle) with the cut off momentum $q_{max}$ hosts the same
	number of states as half of the rhombus at $\bK$ valley.}
	\lb{fig2}
\end{figure}

Near the Dirac points, $\bK = (4\pi/3,0)$ or $\bK' = (2\pi/3, 2\pi/\sqrt 3)$, shown in Fig.~\ref{fig2}, the energy
dispersion becomes linear when expressed via relative small differences $\bq = \bk - \bK$ or $\bq = \bk - \bK'$:
\be
\pm\e_\bk=\pm\e_{\bq+\bK^{(\prime)}}\approx\pm\frac {\sqrt 3}2 t q \equiv\pm\e_q,
\lb{disp}
\ee
while the hopping phases in those vicinities mainly follow the azimuthal angle of $\bq$: $\theta_q = \arctan
q_y/q_x$, up to some shift and sign inversion ($\bK$ and $\bK'$ valleys revealing opposite circularities):
\be
 \arg \g_\bk \approx
 \begin{cases}
 \phantom{-}\theta_q + \pi\,, & \bk = \bq + \bK\,,\\
 -\theta_q + \frac{2}{3}\pi\,, & \bk = \bq + \bK'\,.
 \end{cases}
 \lb{phi}
 \ee
This permits us to label the low-energy graphene characteristics by the valley index, and the reduced quasi-momentum
$\bq$ referred to that valley, we reserve the general symbol $\bk$ for the quasi-momentum measured from the BZ center.
From the low-energy point of view, the standard momentum sum over the whole Brillouin zone is conveniently approximated
by the integral over the equivalent circular areas centered at $\bK$ and $\bK'$ valleys:
\be
\begin{aligned}
\frac{1}{N}\sum_\bk f_\bk
 &=\frac{1}{N}\sum_\bq f_{\bK+\bq}+f_{\bK'+\bq}\\
 &=\bigl[\text{assuming: $ f_{\bK+\bq}=F_\bq\, \&\, f_{\bK'+\bq}=G_{\bq}$}\bigr]\\
 &\approx \frac{1}{2\pi}\frac{1}{q_{max}^2}\int\limits_0^{2\pi}
d\theta\hspace{-1mm} \int\limits_0^{q_{max}}\hspace{-1mm} dq\,q \left( F_{\bq}+G_{\bq} \right),
\end{aligned}
\lb{int1}
\ee
with the radius $q_{max} = 2\sqrt{\pi/\sqrt3}$ chosen in a way to preserve the total number of states as in the original
rhombic BZ displayed in Fig.~\ref{fig2}. Then, the linear isotropic approximation of the graphene energy dispersion law,
Eq.~\ref{disp}, can be rewritten as:
\be
\e_q \approx  W \frac{q}{q_{max}},
\lb{dispq}
\ee
where the effective graphene bandwidth
\be
W = (\sqrt3/2)tq_{max} = \sqrt{\pi\sqrt3}\,t
\ee
is somewhat reduced compared to the real bandwidth value $3t$.

At sufficiently low temperatures $T$, electronic dynamics of a many-body system is conventionally described by the
(advanced) Green's functions (GF's) \cite{Bonch-Bruevich}, whose Fourier-transform in the energy domain reads:
\be
\llang A|B\rrang_\e = \frac i{\pi} \int_{-\infty}^0  {\rm e}^{i(\e - i0)t}\langle\left\{A(t),
B(0)\right\}\rangle dt.
\lb{gf}
\ee
This involves the grand-canonical statistical average: $\langle O\rangle = {\rm Tr}\,\left[{\rm e}^{-(H - \mu)/
k_{\mathrm{B}}T}O_H(t)\right]\bigl/\,{\rm Tr}\,\left[{\rm e}^{-(H - \mu)/k_{\mathrm{B}}T}\right]$ of an operator $O_H(t)
= {\rm e}^{iHt}O{\rm e}^{-iHt}$ in the Heisenberg representation. Here and below $\{.,.\}$ represents the anticommutator and
$[.,.]$ the commutator of two operators. The GF energy argument $\e$ implicitly includes an infinitesimal negative imaginary
part, as shown explicitly in Eq.~\ref{gf} for the Fourier exponent.

As known \cite{Bonch-Bruevich,Economou}, GF's satisfy the equation of motion:
\be
\e \llang A|B\rrang_\e = \langle\left\{A(0),B(0)\right\}\rangle + \llang [A,H]|B\rrang_\e.
\lb{de}
\ee
For practical reasons, in what follows the energy sub-index at GF's is either omitted, or enters directly as an argument.

A convenient description of the two-band graphene system, Eq.~\ref{Hb}, is given in terms of 2$\times2$ GF matrices (in
conduction and valence bands indices): $\hat G_{\bk,\bk'} = \langle\langle \psi_\bk^{\phantom{\dagger}}|\psi_{\bk'}^\dagger
\rangle\rangle$, based on the band operators arranged in (column and row) spinors:
\be
 \psi_\bk = \left(\begin{array}{c}\b_{+,\bk}\\\b_{-,\bk} \end{array}\right),\qquad
  \psi_\bk^\dagger = \left(\b_{+,\bk}^\dagger,\b_{-,\bk}^\dagger\right).
   \lb{sp}
\ee

Knowledge of GF's permits to obtain, in principle, all the observables of the system. For instance, the density of states
(DOS) is expressed as:
\be
\r(\e) = \frac1\pi{\rm Im\,Tr\,}\hat G_{loc},
\lb{rho}
\ee
via the locator GF matrix:
\be
\hat G_{loc} = \frac1N \sum_\bk \hat G_\bk\,,
\lb{gloc}
\ee
involving the momentum-diagonal GF matrices $\hat G_{\bk,\bk} \equiv \hat G_\bk$. Then the Fermi level $\e_{\rm F}$ in
the electronic spectrum is defined by the equation:
\be
\int_{-\infty}^{\e_{\rm F}} \r(\e) d\e = Q,
\lb{eF}
\ee
where $Q$ is the number of charge carriers per unit cell.

In absence of impurities, the exact solution for GF matrices is: $\hat G_{\bk,\bk'} = \d_{\bk,\bk'}\hat G_\bk^{(0)}$,
where the non-perturbed momentum-diagonal GF:
\be
\hat G_\bk^{(0)}(\e) = \frac{\e\,\hat 1 + \e_\bk\,\hat\s_3}{\e^2 - \e_\bk^2}
\lb{g0}
\ee
includes the identity $\hat 1$ and the 3rd Pauli matrix $\hat\s_3$ acting in the band space. Then the explicit locator
matrix is found with the help of Eq.~\ref{int1} as:
\be
\hat G_{loc}^{(0)}(\e) \approx \frac{2}{W^2}\left(\begin{array}{cc}
     -W + \e\ln\frac{\e}{\e - W} & 0  \\
    0 & W + \e\ln\frac{\e}{\e + W}
\end{array}\right),
\lb{G0}\ee
and defines the corresponding DOS per graphene unit cell, $\r_0(\e) = 2\pi^{-1}{\rm Im\,}G^{(0)}(\e)$, where we denoted
\be
\frac12{\rm Tr\,}\hat G_{loc}^{(0)}(\e) \equiv G^{(0)}(\e) = -\frac\e{W^2}\,\ln\left(1 - \frac{W^2}{\e^2}\right).
\lb{fe}
\ee
This results in the known linear DOS at low energies:
\be
\r_0(\e) \approx \frac {2|\e|}{W^2}\,\Theta\left(W^2 - \e^2\right),
\lb{re}
\ee
with the Heaviside step function $\Theta(x)$.

Then, considering $Q = 1$ in Eq.~\ref{eF}, the unperturbed Fermi level locates just at the Dirac point: $\e_{\rm F} = 0$,
but it would be displaced under impurity effects modifying both $\r(\e)$ and $Q$.

\section{Impurity effects in Lifshitz model}
\lb{Lif}
To study the impurity effects in graphene, we build the perturbation Hamiltonian in analogy with the well studied models
in the theory of disordered solids. In what follows we consider two such models: the Lifshitz isotopic model (LM) \cite{Lifshitz1},
most adequate for substitutional impurities, and the Anderson $s$-$d$ hybrid model (AM) \cite{Anderson1}, suitable for interstitial
or adatom impurities.

Let us begin from the simpler LM case where impurities are supposed to substitute host carbon atoms at random sites $\br_j$
($j$ stands for type/sublattice), and the impurity Hamiltonian contains a single perturbation parameter, $V$, the on-site
energy difference between the impurity and host atomic levels. Such Hamiltonian is presented in terms of local operators:
\be
H_{\rm LM} = V \sum_{\br_j} b_{\br_j}^\dagger b_{\br_j}^{\phantom{\dagger}},
\lb{HL}
\ee
and a GF treatment of this LM perturbation on graphene spectrum was recently discussed \cite{Skrypnyk2018} and here we
shall consider it only to compare with the alternative AM situation. So, rewriting Eq.~\ref{HL} in terms of $\psi$-spinors,
Eq.~\ref{sp}, it permits to generalize the ordinary single-band scattering:
\be
H_{\rm LM} = \frac 1{2N} \sum_{\br_j,\bk,\bk'} \psi_\bk^\dagger\,  \hat V_{\br_j,\bk,\bk'}^{\phantom{\dagger}}\,
\psi_{\bk'}^{\phantom{\dagger}}.
\lb{HLs}
\ee
Here the scattering matrices:
\be
\hat V_{\br_j,\bk,\bk'} = 2V\exp\left[i(\phi_{\br_j,\bk} - \phi_{\br_j,\bk'})\right]\hat m_j
\lb{sc}
\ee
contain the matrix kernels:
\be
\hat m_j = \frac12\left[\hat 1 - (-1)^j\hat \s_1\right].
\lb{mk}
\ee
The latter include both the intra-band scattering processes (unit matrix) and the inter-band ones (Pauli $\hat\s_1$
matrix) and form an idempotent and normalized matrix algebra:
\be
\hat m_j\hat m_{j'} = \d_{j,j'}\hat m_j,\ \
\hat m_1 + \hat m_2 = \hat 1,\ \
\hat m_1 - \hat m_2 = \hat \s_1.
\lb{ma}
\ee

The most relevant GF under impurity scattering is the momentum-diagonal part, $\hat G_{\bk,\bk} \equiv \hat G_\bk$,
which gets modified from Eq.~\ref{g0} to:
\be
\hat G_\bk^{-1} = \left(\hat G_\bk^{(0)}\right)^{-1} -\ \hat \S_\bk^{\phantom{-1}},
\lb{se}
\ee
where the self-energy $\hat \S_\bk$ is also a matrix in the band space. It can be generally expressed through the so
called group expansion (GE) \cite{Lifshitz,Ivanov1987,Loktev2015}, a series in powers of impurity concentration $c$
(defined as the number of impurities per host site):
\be
\hat \S_\bk = c \hat T_\bk\left(1 + c\hat B_\bk + \dots\right).
\lb{ge}
\ee
Here, the T-matrix, $\hat T_\bk$, takes into account all multiple scatterings of the $\bk$-th band state on the same
impurity center while the terms in parentheses next to unity result from all such scatterings on clusters of two,
$\hat B_\bk$, and more impurity centers. The detailed structure of $\hat B_\bk$ is presented in what follows, considering
an onset of cluster dominated scattering.

In the simplest case when all the GE terms in Eq.~\ref{ge} besides unity can be neglected, the T-matrix approximation,
$\hat \S_\bk \approx c \hat T_\bk$, dominates. For the system with Hamiltonian $H_0 + H_{\rm LM}$, there are two partial
contributions into the total T-matrix, each labeled by the index $j$ that specifies sublattice position of an impurity site
$\br_j$. Those partial $\hat T_{j}$'s are expressed through the scattering matrices $\hat V_{\br_j,\bk,\bk'}$, Eq.~\ref{sc},
via the multiple scattering series:
\bea
\hat T_{j,\bk} & \equiv & \hat T_{\br_j,\bk} = \hat V_{\br_j,\bk,\bk}\nn\\
& + & \frac 1{2N} \sum_{\bk'}\hat V_{\br_j,\bk,\bk'}\,\hat G_{\bk'}^{(0)}\,\hat V_{\br_j,\bk',\bk} + \dots.
\lb{tmat}
\eea
Since all the phase factors ${\rm e}^{i\phi_{\br_j,\bk}}$ get fully compensated here, $\hat T_{j,\bk}$ result to be
momentum independent, $\hat T_{j,\bk} \to T(\e)\hat m_j$, with the energy-dependent scalar factor:
\be
 T(\e) = \frac{V}{1 - VG^{(0)}(\e)}.
\lb{tj}
\ee
Moreover, the idempotency of $\hat m$'s, Eq.~\ref{ma}, implies that the total self-energy is summed up to $c\hat T(\e) =
T(\e)\left(c_1\hat m_1 + c_2\hat m_2\right)$, where $c_j$ is the partial impurity concentration on $j$-th sublattice.

As usual in LM, and also in the analogous models~\cite{Ivanov1987,Loktev2015}, the impurity resonance $\e_{res}$ is defined
by the $T(\e)$ pole, in our case this resonance condition reads:
\be
V{\rm Re\,}G^{(0)}(\e_{res}) = 1,
\lb{res}
\ee
and from the explicit result by
\ref{fe} it is readily found that the condition by Eq.~\ref{res}, can be only reached for quite a strong perturbation: $|V| \geq 1.44W \approx 9$ eV (here and in what follows we use the commonly adopted value of $t =
2.6$ eV). Even though the unitary limit of infinitely strong perturbation, $V \to \infty$, is commonly used to describe the
zero energy resonance by vacancies in graphene \cite{Pereira2006,Pereira2008}, the above $V$ strength seems quite unrealistic
for substitutional impurities in graphene, especially for carbon near neighbors in the periodic table. Since in this case one
expects $|V| \ll 9$\,eV, the T-matrix denominator in Eq.~\ref{tj} can be approximated to unity (neglecting also its small
imaginary part), then one recovers the Born approximation result:
\bea
\hat \S_\bk & \approx & c\hat T \approx 2V(c_1\hat{m}_1 + c_2\hat{m}_2)\nn\\
& = &  V \bigl[(c_1 + c_2)\hat 1 + (c_1 - c_2)\hat\s_1\bigr].
\eea
Even in this simplest Born limit, the resulting spectrum strongly depends on the partial impurity occupations of two
graphene sublattices. Defining the total impurity concentration $c = c_1 + c_2$, and the sublattice impurity occupation
asymmetry $\D c = c_1 - c_2$, the spectral Eq.~\ref{se} takes the explicit form:
\be
\hat G_\bk^{-1}(\e) = \bigl(\e - c V \bigr)\hat 1 - \e_q\hat \s_3 - \D c V\hat\s_1\,,
\lb{seLM}
\ee
and provide the restructured energy dispersion relations (by the poles of $\hat G_\bk$):
\be
\E_{\pm,\bk} \approx c V \pm \sqrt{\e_q^2 + \left(\D c V\right)^2},
\lb{LMRS}
\ee
\begin{figure}
\centering
\includegraphics[scale=0.6]{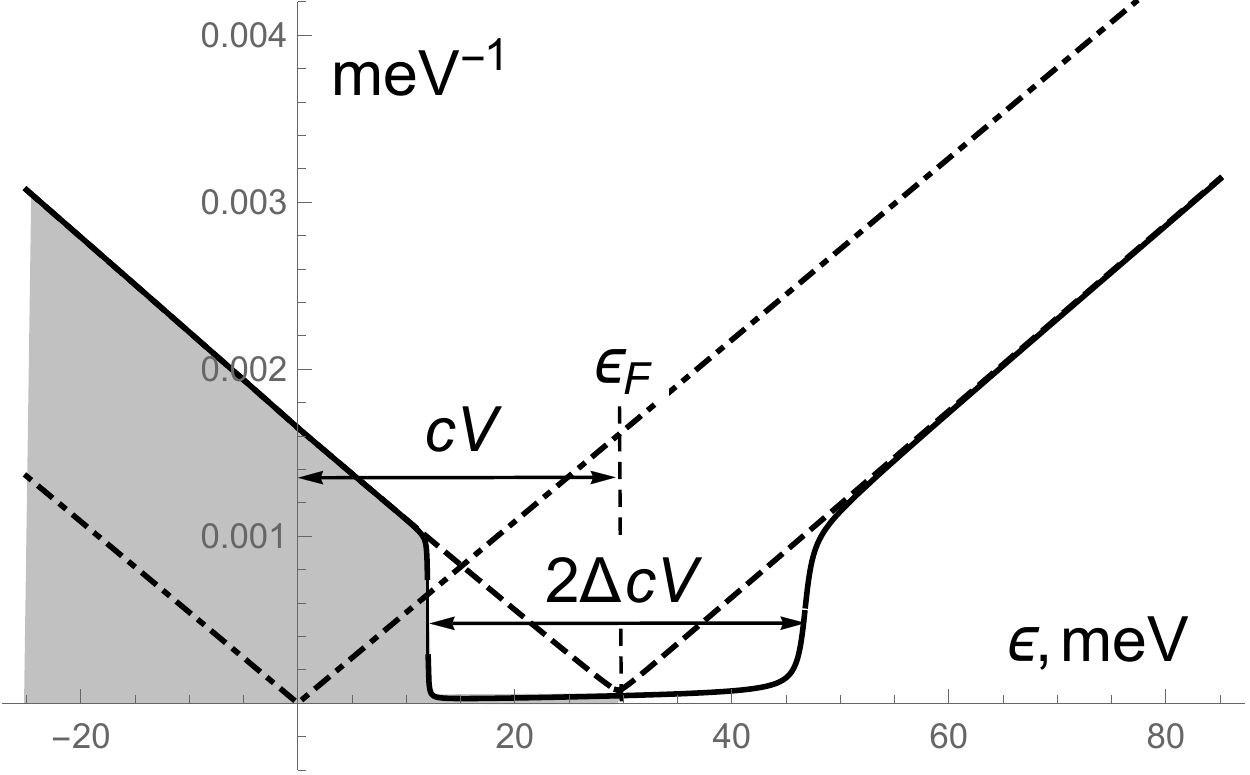}%Fig3
\caption{Linear low energy DOS of pure graphene (dash-dotted line), and its restructured DOS (solid line) under LM impurities
with on-site disorder strength $V = W/2$ and concentration $c = 10^{-2}$ that presents: 1) the global energy shift by $cV$ and
2) the spectrum gap of $2\D c|V|$ around the shifted Fermi level (dashed line), due to the asymmetry of impurity occupation of
host sublattices, $\D c = 6\cdot 10^{-3}$. The range of filled quasiparticle states is shadowed.}
\lb{fig3}
\end{figure}

In the most natural case of equal sublattice occupancies, $\D c = 0$, the inter-band scattering (the $\hat\s_1$-term)
cancels out and Eq.~\ref{LMRS} takes particularly simple form. The overall impurity effect gets reduced just to a simple
mean-field shift of the energy reference by $cV$ with no other notable changes in the observable properties. However,
if there exists a certain occupational asymmetry between the two sublattices, $\D c \neq 0$, for instance due to lattice
buckling, the spectral Eq.~\ref{seLM} would retain also a finite off-diagonal term. As a consequence, apart of the Fermi
level shift $cV$, there appears also a splitting of the valence and conduction bands quantified by a finite gap value
$2\D c|V|$, see Eq.~\ref{LMRS}. This would, respectively, modify the low energy DOS, and the corresponding gapped-like
analog of Eq.~\ref{re} reads:
\be
\r(\e) \approx \frac{2|\e - c V|}{W^2}\,\Theta\left[(\e - c V)^2 - (\D c V)^2\right]\,,
\lb{rem}
  \ee
recovering purely linear behavior beyond the gap, unlike peculiar behaviors near impurity resonances in AM (see in detail
in the next sections). Validity of this simplest Born approximation picture is also confirmed by the full T-matrix calculation
for DOS at the choice of $V = W/2$, $c = 10^{-2}$ and $\D c = 6\cdot 10^{-3}$, displayed in Fig.~\ref{fig3}.

\section{Anderson's impurity model, a general discussion}
\lb{And}
Anderson model (AM) differs from the Lifshitz one by considering impurities beyond the host sites, for instance, impurity
adatoms over the graphene plane. The model introduces new degrees of freedom into the system by means of impurity Fermi
operators $c_\br$. We label them by in-plane projection vectors $\br$ that are not necessarily lattice sites, and hence
not bearing the sublattice index. Another specifics of AM is the dynamics of impurity perturbation, which is described
by two independent parameters; the impurity energy level (on-site energy) $\e_0$, and the hopping (coupling) parameter
$\o$ of its hybridization with carbons at nearest neighbor graphene sites $\bn_j$. In terms of local operators, this
perturbation Hamiltonian reads:
\be
H_{\rm AM} = \sum_\br \left[\e_0 c_\br^\dagger c_\br^{\phantom{\dagger}} + \o \sum_{\langle \br, \bn_j\rangle}
\left(b_{\bn_j}^{\dagger}c_\br^{\phantom{\dagger}} + h.c.\right)\right].
\lb{HA}
\ee
Also a GF treatment of this perturbation was proposed previously \cite{Skrypnyk2013} and here we shall develop it in
a more general context. Thus, expressing again the local atomic operators $b_{\bn_j}^\dagger$ through the graphene band
$\psi^\dagger_\bk$ spinors, Eq.~\ref{sp}, the above Hamiltonian is brought to the form:
\bea
H_{\rm AM} & = & \sum_{\br} \left[\e_0 c_{\br}^\dagger c_{\br}^{\phantom{\dagger}}\right.\nn\\
& + & \left.\frac{\o}{\sqrt {N}} \sum_\bk \left(\psi_\bk^\dagger\,u_{\bk,\br}^{\phantom{\dagger}}\,
 c_{\br}^{\phantom{\dagger}} + h.c.\right)\right],
\lb{HAb}
\eea
where the form-factor (column) spinor $u_{\bk,\br}$ reflects the local symmetry of an impurity at position $\br$
and is given as:
\be
u_{\bk,\br} = \frac 1{\sqrt 2} \sum_{\langle \br, \bn_j\rangle} {\rm e}^{i\phi_{\bn_j,\bk}}
\left(\begin{array}{c} 1\\(-1)^{j - 1}\end{array}\right),
\lb{ss}
\ee
with the hopping phases $\phi_{\bn_j,\bk}$ by Eq.~\ref{Ang}.

Considering the equation of motion for the momentum-diagonal GF matrix we have:
 \be
 \hat G_\bk = \hat G_\bk^{(0)} + \frac{\o}{\sqrt{N}}\sum_{\br} \hat G_\bk^{(0)}\,
 u_{\bk,\br}^{\phantom{\dagger}}\,\llang c_\br^{\phantom{\dagger}}|\psi_\bk^\dagger\rrang,
 \lb{eq1}
 \ee
where the impurity-host GF (forming a row spinor in band indices), $\llang c_\br^{\phantom{\dagger}}|
\psi_\bk^\dagger\rrang$, can be excluded from that equation using its own equation of motion:
 \be
 \llang c_\br^{\phantom{\dagger}}|\psi_\bk^\dagger\rrang = \frac{\o}{\left(\e - \e_0\right)\sqrt{N}} \sum_{\bk'}
 u_{\br,\bk'}^\dagger\, \hat G_{\bk',\bk}^{\phantom{\dagger}},.
 \lb{eq2}
 \ee
This effectively decouples host-host and impurity-host GF's to give:
\be
 \hat G_\bk = \hat G_\bk^{(0)} + \frac{1}{N}\sum_{\br,\bk'} \hat G_\bk^{(0)}\, \hat{V}_{\br,\bk,\bk'}
 ^{\phantom{\dagger}}\,\hat G_{\bk',\bk}\,,
 \lb{eq2}
 \ee
where the effective $2\times 2$ scattering matrix (in the band indices) for the impurity at $\br$ position reads:
\be
 \hat V_{\br,\bk,\bk'}^{\phantom{\dagger}} = u_{\bk,\br}^{\phantom{\dagger}}\,\frac{\o^2}{\e - \e_0}\,
 u_{\bk',\br}^\dagger\,.
 \lb{vef}
\ee
The detailed structure of the $\hat V_{\br,\bk,\bk'}$ matrices follows from the particular $j$-types of graphene
sites $\bn_j$, neighbors to $\br$, as in Eq.~\ref{ss}.

Despite the AM scattering matrix, Eq.~\ref{vef}, differs from the former LM one, Eq.~\ref{sc}, by its explicit energy
dependence, it generates formally the same GE series in powers of $c$ as the LM result. For a general scattering problem,
the momentum diagonal T-matrix with the corresponding $u_{\bk,\br}$ spinor reads:
\be
\hat{T}_{\br,\bk}(\e) = \frac{\o^2 u_{\bk,\br}^{\phantom{\dagger}}\,u_{\bk,\br}^{{\dagger}}}{\e - \e_0 - \o^2N^{-1}\,
\sum_{\bk'}u_{\bk',\br}^{{\dagger}}\,\hat G_{\bk'}^{(0)}(\e)\,u_{\bk',\br}^{\phantom{\dagger}}},
\lb{ttk}
\ee
formally the same as for the LM scenario, consult Eqs.~\ref{tmat}~and~\ref{tj}.

\begin{figure}
	\centering
	\includegraphics[scale=0.4]{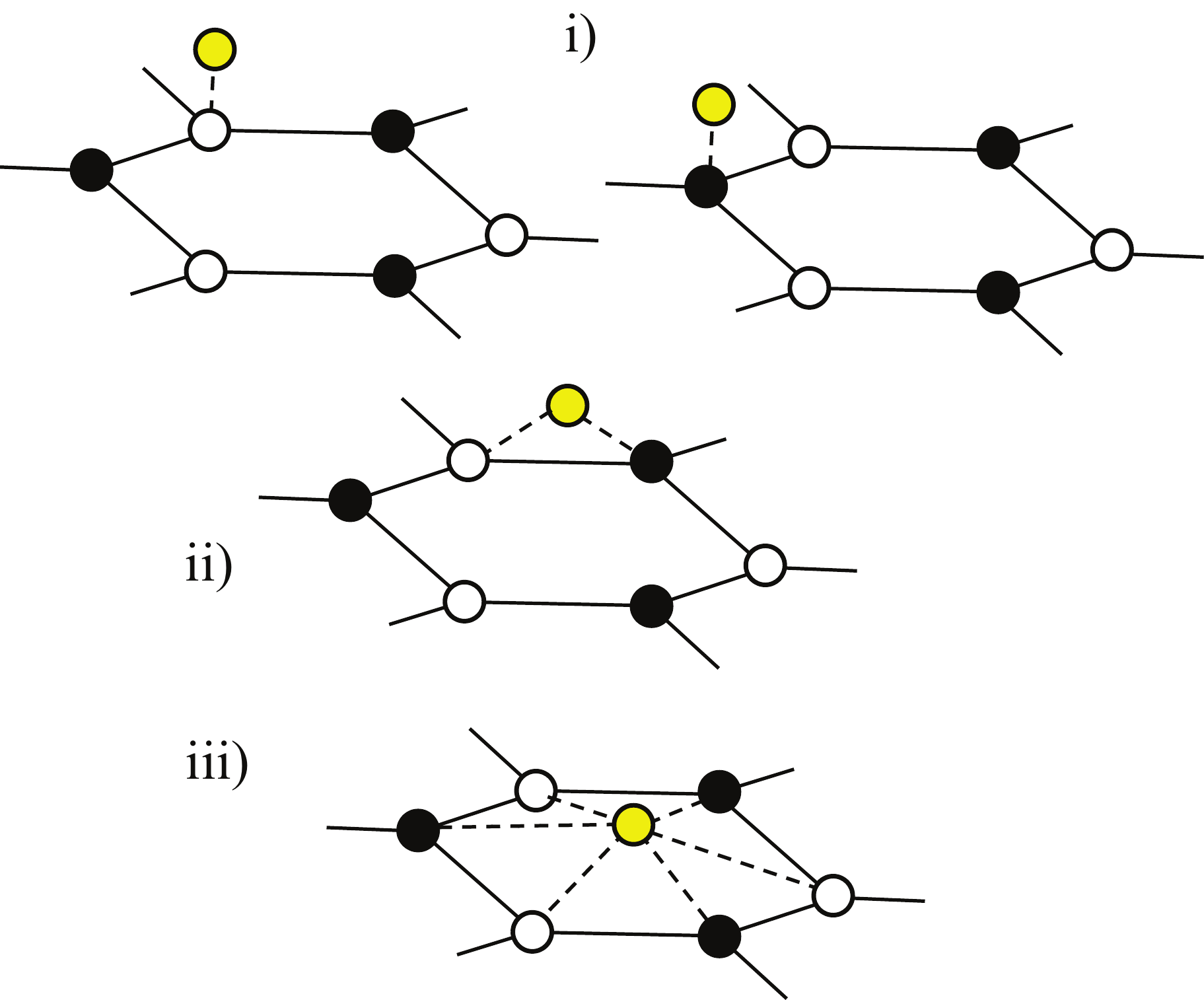}%Fig4
	\caption{Different positions of AM impurities (yellow circles) over a graphene layer: i) t-positions, atop of carbon
	lattice sites of 1- or 2- types, ii) b-positions, over centers of bridges between 1- and 2-type lattice sites (one of
	three possible bridges shown), iii) h-position, over the center of a hexagonal cell.}
	\lb{fig4}
\end{figure}

 The most natural positions discussed in what follows are those shown in Fig.~\ref{fig4} and categorized as:

i) top position (t-position), impurity projects just on a host lattice site $\bn_j$ and such position can be indexed
by this $j$,

ii) bridge position (b-position), impurity projects on a midpoint $\br$ between two carbons belonging to the opposite
sublattices. In this case the positions of two hybridizing carbons are: $\bn_{1,i} = \br - \boldsymbol\d_i/2$, and
$\bn_{2,i} = \br + \boldsymbol\d_i/2$, where three nearest-neighbor vectors $\boldsymbol\d_{i=1,2,3}$ are displayed in
Fig.~\ref{fig1}. The corresponding bridge configurations b$_{\boldsymbol \d_i}$ are related through $\pm 120^\circ$
rotations.

iii) hollow position (h-position), impurity projects on a center of hexagonal lattice cell, in this case we have three
nearest neighbor sites $\bn_{1,i}= \br + \boldsymbol\d_i$, $i=1,2,3$ from the sublattice 1, and three such sites $\bn_{2,i}
= \br - \boldsymbol\d_i$ from the sublattice 2.

So, generally, there are two possible types of t-position (t$_j$), three types of b-position (b$_{\boldsymbol \d}$), and
a single type of h-position. Obviously, two t$_j$-types can be occupied either symmetrically or asymmetrically in $j$,
while such occupations of three b$_{\boldsymbol \d}$-types and of single h-type are $j$-independent. A special difference
between them is yet in possible momentum dependence for the self-energy and T-matrix (besides their common $\e$ dependence).
This effect is especially pronounced in the h-case, making it qualitatively different from the t- and b-cases. It can be
also shown that, due to their different couplings to the graphene host, the listed three positions will contribute into the
system dynamics in different energy ranges, and therefore they can be considered independently.

\begin{table}[h!]
  \begin{center}
    \caption{AM tight-binding parameters $\e_0,\o $ for some representative top impurity adatoms on graphene, including the ``gauge'' value $\omega^\ast$ discriminating between weak, strong and intermediate perturbations.}
    \label{tab:table1}
    \begin{tabular}{|c|c|c|c|c|}
    \hline
      Atom & Cu & Cu & H & F \\
      \hline
      Position & t- & b- & t- & t- \\
      \hline
      $\e_0$ (eV) & 0.08 & 0.02 & 0.16 & -2.2 \\
      $\o$ (eV) & 0.81 & 0.54 & 7.5 & 5.5 \\
      $\o^\ast$ (eV) & 1.99 & 1.73 & 2.17 & 4.35\\
      \hline
         \end{tabular}
  \end{center}
\end{table}

Available data suggest that adsorption in the top position seems to be favorable for light atoms like hydrogen~
\cite{Boukhvalov:PRB2008,Gmitra:PRL2013}, fluorine~\cite{Wu:APL2008,Sahin:PRB2011,Irmer2015:PRB} and copper~
\cite{Wu:APL2009,Amft:JPhysCondMat2011,Frank:PRB2017}, the heavier gold atom \cite{Chan2008:PRB,Amft:JPhysCondMat2011},
and, for example, also the light ad-molecule methyl \cite{Zollner:Meth2016}. A special case is the vacancy which as
was mentioned induces a zero-energy mode~\cite{Ducastelle:PRB2013,Pereira2006,Peres:PRB2006,Pereira2008}.

In the following sections we consider in more detail each of the above mentioned impurity positions, and analyze
reconstructed spectra and localization properties of the corresponding eigenstates. This will be illustrated for
several particular examples of impurity adatoms whose known AM parameters are collected in Table 1.

\section{Anderson's impurities at top position}
\lb{pos}

For a t-position impurity located at $\br_j$, the form-factor spinor, Eq.~\ref{ss}, is realized as:
\be
u_{\br_j,\bk} = \frac 1{\sqrt 2}{\rm e}^{i\phi_{\br_j,\bk}}\left(\begin{array}{c} 1\\(-1)^{j-1}\end{array}\right),
\lb{upk}
\ee
and the corresponding effective scattering matrix then reads:
\be
\hat V_{\br_j,\bk,\bk'} = \frac{\o^2}{\e - \e_0}\exp\left[i(\phi_{\br_j,\bk} - \phi_{\br_j,\bk'})\right]\hat m_j
\lb{veft}
\ee
with the same $\hat m_j$ matrices as in the LM case, see Eq.~\ref{mk}. Defining the energy dependent effective
scattering potential:
\[V(\e) = \frac{\o^2}{\e - \e_0},\]
the corresponding T-matrix in AM takes an analogous form to the LM case, Eq.~\ref{tj}: $\hat T_{\br_j,\bk} = T_t(\e)
\hat{m}_j$, where the scalar T-factor:
\be
T_t(\e) = \frac{V(\e)}{1-V(\e)G^{(0)}(\e)} = \frac{\o^2}{\e - \e_0 - \o^2 G^{(0)}(\e)}\,,
\lb{tm}
\ee
is, alike the LM case, momentum and sublattice independent.

The condition for impurity resonances, the real part of T-matrix denominator becoming zero, leads here to the
explicit equation:
\be
\e_{res}\left(1 + \frac{\o^2}{W^2} \ln \frac{W^2 - \e_{res}^2}{\e_{res}^2}\right) = \e_0.
\lb{res0}
\ee
Comparing to the LM case, Eq.~\ref{res}, there are no special restrictions on AM perturbation parameters for such
resonance to appear. It is a matter of fact that the hybridization $\o$ between the adatom and graphene host is
responsible for the shifts of the resonance energy, $\e_{res}$, towards zero, when comparing with the initial atomic
level $\e_0$ (supposing the latter satisfies $\e_0^2 < W^2/2$). The relative magnitude of this shift depends on
the coupling parameter $\o$ compared to its ``gauge'' value:
\be
\o^\ast = W/\sqrt{\ln\left(W^2/\e_0^2 - 1\right)}.
\ee
This distinguishes between the three coupling types:

(a) \emph{weak}, $|\e_{res} - \e_0| \ll |\e_0|$, for $|\o| \ll \o^\ast$,

(b) \emph{strong}, $|\e_{res}| \ll |\e_0|$, for $|\o| \gg \o^\ast$, and

(c) \emph{intermediate}, $|\e_{res} - \e_0| \sim |\e_0|$, for $|\o| \sim \o^\ast$.

Then, from the comparison of $\o$ to $\o^\ast$ for the cases in Table \ref{tab:table1}, Cu adatoms at t- and b-positions
can be classified as weakly coupled, H adatoms at t-position as strongly coupled, and F adatoms at t-position as
intermediate coupled.

In particular, for weakly coupled impurities, the approximate solution of Eq.~\ref{res0} is given within to logarithmic
accuracy as:
\be
\e_{res} \approx \frac{\e_0}{1 + (\o/\o^\ast)^2}.
\lb{res1}
\ee

Our next studies consider the band structure reconstruction for symmetric and asymmetric sublattice occupancies,
and the arise of mobility edges for t-positioned AM impurities. The starting point for those discussions is the
spectral equation for the inverse of momentum-diagonal GF matrix, $\hat G_\bk^{-1}(\e)$. In analogy with the LM,
Eq.~\ref{seLM}, the T-matrix approximation averaged in disorder by t-position AM impurities reads here:
\be
\hat G_\bk^{-1}(\e) = \bigl[\e - c T_t(\e)\bigr]\hat 1 - \e_q\hat \s_3 - \D c T_t(\e) \hat\s_1\,.
\lb{seAMtop}
\ee
The restructured band spectrum in presence of impurities is usually sought as the roots of secular equation
\cite{Bonch-Bruevich}:
\be
{\rm Det}\,\hat G_\bk^{-1}(\e) = 0.
\lb{chaAMtop}
\ee
In fact, this is an essential reduction of the underlying eigenvalue problem for the full, translationally non-invariant
Hamiltonian $H_0 + H_{\rm AM}$ with randomly disordered impurities \cite{Lifshitz} that intrinsically admit alternation
of the band-like and localized ranges, the celebrated \emph{metal-insulator transitions} \cite{Mott1967}. The above
secular Eq.~\ref{chaAMtop} with use of Eq.~\ref{seAMtop} provides just a disorder averaged approximation where the
quasi-momentum $\bk$ is no more an exact quantum number as it was for the unperturbed band spectrum, e.g.~in Eq.~\ref{Hb}.

One way how to construct solutions of the secular Eq.~\ref{chaAMtop} is to look at energy {\it vs} quasi-momentum relation,
we call it \emph{energy-projected solution} (EPS). Here, for a given real $\bk$, hybridization of each initial $\pm\e_q$
subband with the impurity resonance level $\e_{res}$ generates up to four complex energy roots of Eq.~\ref{chaAMtop}:
$\e = E_{j,\bk} + i\G_{j,\bk}$, $j = 1,\dots,4$. Their real parts $E_{j,\bk}$ approximate the restructured dispersion
laws (for the band-like energy ranges, see also discussion later), while the imaginary parts do the lifetimes $\t_{j,\bk}
\sim \hbar/\G_{j,\bk}$ of such quasiparticles. However, a complicated functional form of $T_t(\e)$, Eq.~\ref{tm}, especially
of the locator GF in its denominator, makes analytical finding of EPS a formidable task. Therefore, some simplifications
are often employed. For example, one identifies restructured energies $E_{j,\bk}$ just with the solutions of the real part of
Eq.~\ref{chaAMtop}:
\be
{\rm Re}\bigl[{\rm Det}\,\hat G_\bk^{-1}(\e)\bigr] = 0\,,
\lb{chaAMtopRe}
\ee
or, goes even simpler, and moves the real part operation deeper into the expression. Particularly, from the determinant to
the self-energy matrix:
\be
{\rm Re}\bigl[{\rm Det}\,\hat G_\bk^{-1}\bigr] \to {\rm Det}\bigl[(\hat G_\bk^{(0)})^{-1} - {\rm Re}\,\hat \S_\bk\bigr]
\ee
or even further just to its denominator:
\be
{\rm Re}\,\hat \S_\bk \to \frac{\o^2 \sum_j c_j \hat m_j }{\e - \e_0 - \o^2{\rm Re}\,G^{(0)}(\e)}.
\lb{den}
\ee
Then, linearizing ${\rm Re}\,G^{(0)}(\e)$ in $\e$ around the resonance allows to find the restructured energies $E_{j,\bk}$
as functions of quasi-momenta $\bk$ in a relatively simple and closed form.

However, there is an alternative way to search for the band-like solutions of Eq.~\ref{chaAMtop} in so-called inverted form,
i.e.~looking for functional dependence of quasimomenta in terms of energy: $\bk(\e)$. Such (in principle complex) solution we call
the \emph{momentum-projected solution} (MPS). In the present case, even keeping the full T-matrix form, the resulting equation turns
to be just an algebraic equation (at most of cubic order) for $\bk(\e)$ or, more precisely, for $q(\e,\theta)$, where $\theta$ stands
for the azimuthal angle of the quasimomentum $\bq=\bk-\bK^{(\prime)}$ (measured relative to the Dirac point).
In the isotropic case, Eqs.~\ref{disp}, \ref{dispq}, one gets the radial component $q$ as a function of $\e$ only. It is obvious that presence
of T-matrix imaginary part (relevant for damping effects) makes this $q(\e)$ generally complex-valued.

Thus, for t-impurities we obtain the MPS explicitly as:
\be
q(\e) = \frac{q_{max}}W \sqrt{\left[\e - cT_t(\e)\right]^2 - \left[\D c T_t(\e)\right]^2}\,,
\lb{qtop}
\ee
with the full complex form of $T_t(\e)$ given by Eq.~\ref{tm}. Another notable advantage of this solution is in providing
a {\it single-valued} $q(\e)$ function, instead of four EPS functions.

Both indicated types of spectral solutions, EPS and MPS, are employed in the following analysis of different AM impurity
cases.

\subsection{Weakly coupled AM t-impurities with symmetric occupancy}

Beginning from the symmetric case, $c_1 = c_2 = c/2$ and $\D c = 0$, one has the inverse GF matrix, Eq.~\ref{seAMtop},
purely diagonal in the sublattice indices, and so the secular equation, Eq.~\ref{chaAMtopRe}, factorizes:
\be
\mathrm{Re}\left[\left(\e - \e_q - c T_t(\e)\right)\left(\e + \e_\bk - c T_t(\e)\right)\right] = 0\,.
\lb{chaAMtop1}
\ee
The above suggested linearization of $T_t(\e)$ denominator, brings this function to the form:
\be
T_t(\e) \approx \frac{\tilde \o^2}{\e - \e_{res} - i\G(\e)},
\lb{tma}
\ee
where the renormalized hybridization strength $\tilde\o$ and the damping term $\G(\e)$ read:
\be
\tilde \o^2 = \o^2\e_{res}/\e_0,\ \ \ \ \G(\e) = \pi|\e|(\tilde \o/W)^2\,.
\lb{tom}
\ee
For weakly coupled AM t-impurities, such linearization is well justified over the whole low-energy range (except for
extremely low values, $|\e| \lesssim W{\rm e}^{-(W/2\tilde\o)^2}$, the latter being as small as $\approx 0.5 \m$eV for
the Cu t-case).

Then, in neglect of damping in Eq.~\ref{tma}, justified for energies not too close to the resonance, $|\e - \e_{res}| \gg
\G(\e_{res})$, the factors in Eq.~\ref{chaAMtop1} provide two decoupled quadratic equations for $\e$. The resulting EPS's
define the explicit low-energy dispersion laws:
\bea
E_{^1_3,\bk} & = & E_{^1_3,\bq+\bK^{(\prime)}} \equiv E_{^1_3,q}\nn\\
 & = & \frac{\e_{res} + \e_q \pm \sqrt{\left(\e_{res} - \e_q\right)^2 + 4 c{\tilde \o}^2}}2\,,
\lb{epm13}		\\
E_{^2_4,\bk} & = & E_{^2_4,\bq+\bK^{(\prime)}} \equiv E_{^2_4,q}\nn\\
 & = & \frac{\e_{res} - \e_q \pm \sqrt{\left(\e_{res} + \e_q\right)^2 + 4 c{\tilde \o}^2}}2\,.
\lb{epm24}
\eea
In the above formulas the subscripts~1,~2~apply to the plus sign, and~3,~4~do to the minus sign. Their validity is
restricted to momenta close to the valleys centers, therefore $\e_q$ can be taken in the linearized form of
Eq.~\ref{disp}.

The restructured energy spectrum around the $\bK$ point for the case of Cu adatoms residing equally on graphene sublattices
with concentration $c = 0.035$ is displayed in Fig.~\ref{fig5}. It illustrates the above mentioned hybridization of two
initial graphene subbands $\pm \e_q$ with the resonance level $\e_{res}$ to produce the energy subbands $E_{j,q}$. Those do
not overlap and fill almost completely the initial spectrum range $(-W,W)$. With growing $c$, the restructured energy
spectrum displays a conjunction of two known scenarios that can take place when a single-band interacts with the impurity
level:
\begin{figure}
	\centering
	\includegraphics[scale=0.6]{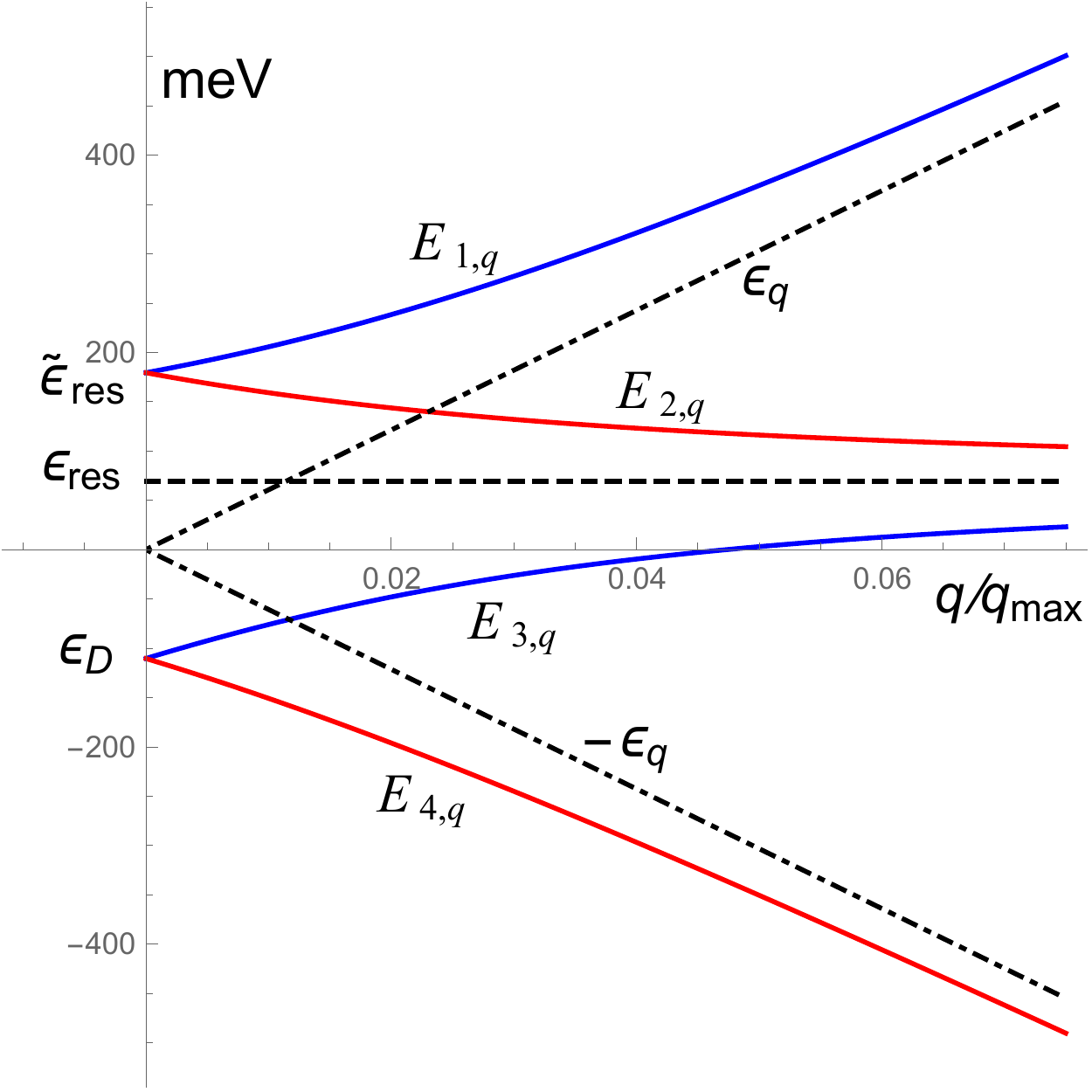}%Fig5
	\caption{Restructured band spectrum (in neglect of its damping) {\it vs} reduced quasi-momentum for graphene with
	Cu t-impurities at concentration $c = 3.5\cdot 10^{-2}$ and symmetric sublattice occupation, Eqs.~\ref{epm13},
	\ref{epm24} (blue and red lines), compared to that for pure graphene, $\pm\e_q$ (dash-dotted lines). The quasi-gap
	between the resonance level $\e_{res} \approx 69$ meV (dashed line) and the bottom of $E_{1,q}$ subband, $\tilde
	\e_{res} \approx 180$ meV, gets filled by the impurity subband $E_{2,q}$ states (see text).}
	\lb{fig5}
\end{figure}

a) Formation of a narrow \emph{quasi-gap} \cite{Ivanov1979} near the resonance level $\e_{res}$ which separates the
branches $E_{1,q}$ and $E_{3,q}$. The quasi-gap exhausts the energy window (assuming $\e_0 > 0$) between ${\rm max}\,
E_{3,q} \approx \e_{res}$ and ${\rm min}\,E_{1,q}$, given by:
\be
\tilde\e_{res} = \e_{res}\frac{1 + \sqrt{1 + c/c^\ast}}2,
\ee
with $c^\ast = \e_0\e_{res}/4\o^2$ (for Cu t-case, $c^\ast \approx 2\cdot 10^{-3}$). Until $c \ll c^\ast$, the quasi-gap
width grows linearly: $\approx\,\e_{res}c/4c^\ast$, then slowing down to $\approx\,\e_{res}(\sqrt{c/c^\ast} - 1)/2$ at $c
\gg c^\ast$. Generally, this results from a strong enough mixing between the intersecting $\e_q$ band and $\e_{res}$ level
(anti-crossing).

b) Formation of a narrow \emph{impurity subband} \cite{Ivanov1987} near the localized level, the $E_{2,q}$ branch that
fills the above indicated quasi-gap, and of a detached weakly affected valence band $E_{4,q}$. The explanation of
that is also very intuitive; the impurity level lies far from the graphene valence band $-\e_q < 0$, and, due to the
weakness of their interaction, both just slightly modify their dispersions ($E_{2,q}$ staying almost dispersionless
and $E_{4,q}$ almost aligned with the original $-\e_q$).

Noteworthy, in the symmetric case ($\Delta c = 0$), the a-type quasi-gap gets completely filled with the states from
the b-type impurity subband, though this filling turns incomplete for an asymmetric occupancy ($\Delta c \neq 0$).

Technically, when considering the full complex T-matrix (either linearized or exact), analytic derivation of EPS from
Eq.~\ref{chaAMtop1} may turn complicated. On the other hand, the MPS, see Eq.~\ref{qtop}, is quite simple and does not
require linearization of $T_t(\e)$ or neglect of its damping.

Within the T-matrix approximation, the momentum-diagonal GF can be written in terms of the unperturbed GF, Eq.~\ref{g0},
but with the shifted argument:
\be
\hat G_\bk(\e) = \hat G_\bk^{(0)}\bigl(\e - c T_t(\e)\bigr)\,.
\ee
This facilitates DOS per unit cell in presence of AM impurities, taking also into account their additional degrees of
freedom (by the $c_\br$ operators) so that the total DOS gets composed of two parts:
\be
\r(\e) = \r_{host}(\e) + \r_{imp}(\e),
\lb{total-dos}
\ee
in an extension of the simpler LM case.

The host part, $\r_{host}(\e) = (\pi N)^{-1}\sum_\bk {\rm Im\, Tr}\,\hat G_\bk$, is analogous to Eqs.~\ref{re}, \ref{rem},
but with the variable energy shift:
\be
\r_{host}(\e) = \frac{2}{\pi}{\rm Im}\,G^{(0)}\bigl(\e - c T_t(\e)\bigr),
\lb{tdos}
\ee
\begin{figure}[h]
	\centering
	\includegraphics[scale=0.6]{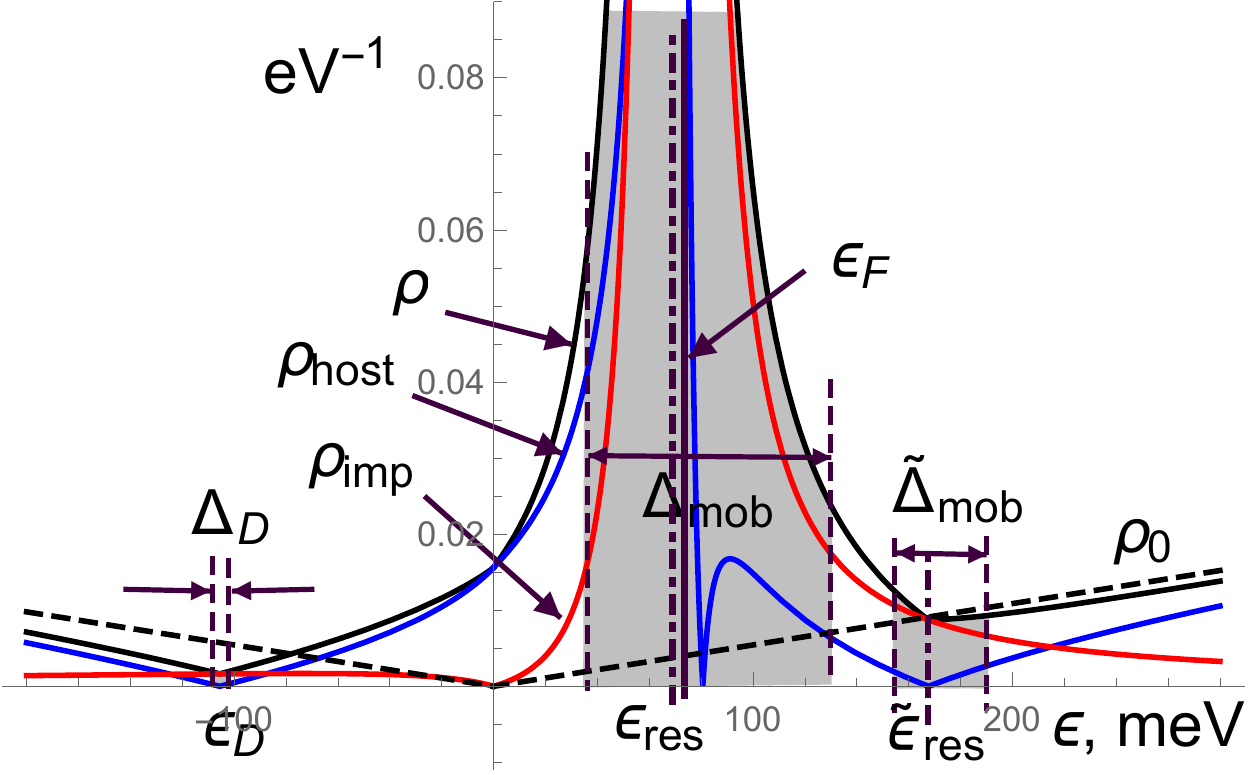}%Fig6
	\caption{Restructured DOS of graphene under Cu impurities as in Fig.~\ref{fig5}, the total value $\r$ (black line)
	and its host, $\r_{host}$ (blue line), and impurity, $\r_{imp}$ (red line), components, referred to the pure graphene
	linear DOS, $\rho_0$ (dashed line). The mobility gaps $\D_{mob}$, $\tilde\D_{mob}$, and $\D_{\rm D}$ (see below)
	are shadowed.}
	\lb{fig6}
\end{figure}

As shown in Fig.~\ref{fig6}, this DOS part displays a sharp peak at $\e_{res}$, and sharp drops towards zero at the quasi-gap
edge, $\tilde\e_{res} = \min{E_{1,\bk}}$, and at $\e_D = \max{E_{4,\bk}} \approx -c\o^2/\e_0$, in consistency with the spectrum
dispersion in Fig.~\ref{fig5}. The last two energies can be seen as ``split Dirac points'': while the ~$\min$~of conduction
band and the~$\max$~of valence band in pure graphene join at the Dirac points, the corresponding $\min$~and~$\max$ of
reconstructed bands in the AM case run off (see also the discussion below).

The impurity DOS part, counting the adatom degrees of freedom, reads:
\be
\r_{imp}(\e) = \frac {1}{\pi} {\rm Im}\,\frac{1}{N}\sum_\br \llang c_\br^{\phantom{\dagger}}|c_\br^\dagger\rrang
\approx \frac{c}{\pi\o^2}\,{\rm Im}\,T_t(\e),
\lb{rimp}
\ee
and, with use of the approximated T-matrix, Eq.~\ref{tma}, it takes the conventional Lorentzian form:
\be
\r_{imp}(\e) \approx \frac{c\e_{res}}{\pi\e_0}\frac{\G(\e)}{\left(\e - \e_{res}\right)^2 + \G^2(\e)}.
\lb{rimp1}
\ee
Comparison of the related contributions to the total $\r(\e)$ in Fig.~\ref{fig6} shows that $\r_{imp}(\e)$ (red line)
generally dominates inside the localization ranges $\D_{mob}$, $\tilde\D_{mob}$, and $\D_{\rm D}$ (see discussion below)
while $\r_{host}(\e)$ dominates outside these ranges.

As already mentioned, the specifics of this band restructuring is the shift of DOS: the zero (Dirac) point moves to
$\e_D = E_{3,0} = E_{4,0} \approx - c\o^2/\e_0$. A fully analogous effect was already met within LM, see the mean-field
shift by $c V$ in Fig.~\ref{fig3}. As a word of caution, the value of $\e_D \approx - c\o^2/\e_0$ lies beyond validity
of the linearized Eq.~\ref{tma}, and was obtained from the exact T-matrix expression, Eq.~\ref{tm}, however, for weakly
coupled impurities, it only slightly differs from $- c\o^2/\e_{res}$ resulting from Eq.~\ref{tma}. This plausibly
justifies the dispersion formulae, Eqs.~\ref{epm13}, \ref{epm24}, for such impurities over the whole low-energy range.

\subsection{Ioffe-Regel-Mott criterium, and mobility gaps}

The presented formal picture of the disorder averaged restructured spectrum at finite concentration of AM impurities
can be considered as consistent and reliable only if the lifetime $\t(E_\bk)$ of the band-like states with quasi-momentum
$\bk$ and energy $E_\bk$ is substantially longer then the intrinsic oscillation period $\l_\bk/v_\bk$ of the associated
Bloch-like wave ($\l_\bk$ being its wavelength and $v_\bk$ the group velocity), i.e.
\be
\frac{\l_\bk}{v_\bk}\ll \t(E_\bk).
\ee
This qualitative and phenomenological
statement is known as the \emph{Ioffe-Regel-Mott} (IRM) \emph{criterion} \cite{Ioffe,Mott1967}. In the simplest case of one
parabolic band centered at the $\G$-point of BZ, the IRM criterion for an extended state with quasi-momentum $\bk$ and
energy $E_\bk$ is commonly written as:
\be
\left.\bk \cdot{\mathbf \nabla}_\bk\, E_\bk\right|_{E_\bk} \gg \hbar\,\t^{-1}(E_\bk),
\lb{irm}
\ee
where one identifies $\l_\bk = 1/|\bk|$ and ${\bf v}_{\bk} = \hbar^{-1}{\mathbf \nabla}_\bk E_\bk$. If, for given $E_\bk$, the lifetime
$\t(E_\bk)$ is too short so that IRM criterion breaks down, and the related state is no more considered as wave-like (or extended), but
localized. Moreover, accordingly to Mott \cite{Mott1967}, if this criterion
fails at least for one $\bk$ on the isoenergetic $E_\bk = \e$ surface (line), then all the states at this energy $\e$ become
localized at impurity centers (or impurity clusters).
Such onset of localization emerges within a certain continuous energy range called the \emph{Mott mobility gap} \cite{Mott1967},
and a threshold between the extended and localized ranges is called the \emph{mobility edge}. One can try to estimate this
edge position by passing from $\gg$ to $\sim$ in Eq.~\ref{irm}, and by using dispersion laws, Eqs.~\ref{epm13}, \ref{epm24},
but taking into account that the used common definition of group velocity and wave length become imprecise near Dirac point,
leaving an uncertainty margin for such procedure. The case of graphene is described below.

At low enough impurity concentrations, the inverse lifetime is well approximated just by the imaginary part of T-matrix,
$\hbar\t^{-1}(\e) = c\,{\rm Im}\,T_t(\e)$, and the latter is given in the vicinity of $\e_{res}$, for example, by the linearized Eq.~\ref{tma}.
That can be used as the right hand side in the IRM criterion for a given $\bk$-state.
The low energy states of graphene have quasi-momenta $\bk$ located near the K-points instead of the $\G$-point and the corresponding Bloch waves
are superpositions of a \emph{standing} $\bK$-wave and \emph{running} $\bq$-waves, but only the latter define the relevant
wavelength scale for the IRM-criterion. Then the product $\bk \cdot{\mathbf \nabla}_\bk$ gets naturally substituted by
$\bq \cdot{\mathbf \nabla}_\bq = q\tfrac\partial{\partial q}$, so that Eq.~\ref{irm} reduces to:
\be
\left|
{q \frac{\partial E_\bq}{\partial q}\Bigr|_{E_\bq}}
\right| \gg c\,{\rm Im}\,T_t(E_\bq) = \hbar\,\t^{-1}(E_\bq).
\lb{irm1}
\ee
This is only half of the story, while taking the momentum derivatives of the EPS dispersion $E_\bq$, Eqs.~\ref{epm13}, \ref{epm24}, is quite impractical.
However, employing MPS, $q(\e,\theta)$, and the reciprocal derivative, $\partial \e/\partial q = \left(\partial q/\partial \e\right)^{-1}$, allow to circumvent
that technical problem and formulate IRM in the equivalent but alternative way:
\be
\left|
\frac{{\rm Re\,}q(\e,\theta)}{\partial {\rm Re\,}q(\e,\theta)/\partial \e}\right| \gg \hbar\,\t^{-1}(\e).
\lb{irmpm}
\ee
Here the relevant wave-number of a Bloch-like wave along angle $\theta$ is represented by ${\rm Re\,}q(\e,\theta)$, the real part
of respective MPS, which can admit anisotropy and that does not require linearized T-matrix. For the considered isotropic t-case,
this corresponds to the real part of Eq.~\ref{qtop}, that can be used in Eq.~\ref{irmpm}. Some more general MPS and the corresponding
mobility edge analyzes will be encountered later.

Let us estimate ranges for IRM to fail, for that we consider the limiting form of Eq.~\ref{irmpm}:
\be
\left|\frac{{\rm Re\,}q(\e)}{\partial {\rm Re\,}q(\e)/\partial\e}\right| \gtrsim \hbar\t^{-1}(\e)
\lb{IRMlim}
\ee
Reaching this limit can be either due to decreasing l.h.s.~of Eq.~\ref{irmpm} or due to growing its r.h.s, and therefore those two cases have different physical
origins. The first case can take place near the split Dirac points, $\tilde\e_{res}$ and $\e_{\rm D}$, where the relevant momenta
tends to go to zero, $q \to 0$, there the analysis can be simplified by using
a {\it linearized in $q$} MPS (LMPS). The second possibility occurs near $\e_{res}$ where the relevant momenta correspond to $\e_q\simeq \e_{res}$ (see Fig.~\ref{fig5}).

Let us estimate for the second case the critical concentration $c_0$, where the IRM breaks down. With growing impurity concentration $c$, the failure of IRM is firstly
expected directly at energy $\e_{res}$, where the inverse lifetime reaches its maximum:
\be
\hbar\,\t^{-1}(\e_{res}) = c\,{\rm Im}\,T_t(\e_{res})= \frac{cW^2}{\pi |\e_{res}|}.
\lb{tmax}
\ee
Contrary, using the simplest LMPS, namely, the unperturbed MPS just for the plain graphene:
$q(\e) \approx q_{max}|\e|/W$, the l.h.s.~of Eq.~\ref{IRMlim} reduces just to $|\e|$. Then, comparing $|\e|$ at resonance energy $\e_{res}$ with
$\hbar\,\t^{-1}(\e_{res})$, we find that IRM inequality holds at $\e = \e_{res}$ until the impurity concentration stays below the critical value:
\be
c \lesssim c_0 = \pi\left(\frac{\e_{res}}W\right)^2.
\lb{c0}
\ee
This just corresponds to the condition that the average distance between neighboring impurities ${\bar r} \sim ac^{-1/2}$
exceeds the resonance state radius $r_{res} \sim a W/\e_{res}$, protecting the coherence of quasi-particles with energies
near $\e_{res}$ (including those near $\tilde\e_{res}$) against random impurity scatterings.

Above this critical concentration, $c > c_0$, the IRM condition breaks down around $\e_{res}$ within a certain finite
energy width $\D_{mob}$, the Mott mobility gap, which gets filled with the localized levels. Using the same unperturbed
LMPS for l.h.s.~of Eq.~\ref{IRMlim} and the Lorentzian form of $\hbar\,\t^{-1}(\e)$ near $\e_{res}$, similar to Eq.~\ref{rimp1},
leads to the estimate:
\be
\D_{mob} \sim \frac{\tilde\o^2}W\sqrt{c - c_0},
\lb{mgap}
\ee
though only valid until $c - c_0 \lesssim c_0$. However, even at $c \gg c_0$ this development can be still traced
analytically. For instance, the result of Eq.~\ref{mgap} stays valid for the lower edge of $\D_{mob}$, only formed
by the states near $\e_{res}$. But for its upper edge, the inverse lifetime $\hbar\,\t^{-1}(\e)$ gets also a growing
contribution from the vicinity of impurity band edge $\tilde\e_{res}$ and the corresponding term, $\tilde\D_{mob}$, can
be estimated with the proper LMPS, $q(\e) \approx q_{max}|\e - \tilde \e_{res}|/W$, used in Eq.~\ref{IRMlim}:
\be
\tilde\D_{mob} \sim \hbar\t^{-1}(\tilde\e_{res} - \tilde\D_{mob}) \sim \frac{c^{1/3}\tilde\o^{4/3}\e_{res}^{1/3}}{W^{2/3}}.
\lb{Dtd}
\ee
The latter value exceeds the impurity band width, $c\tilde\o^2/\e_{res}$, formally defined by Eq.~\ref{epm24}, making this
band unphysical as far as $c \lesssim \sqrt{c_0 c_\ast}$, where $c_\ast = (\e_{res}/\tilde\o)^2$.

\begin{figure}[h]
	\centering
	\includegraphics[scale=0.6]{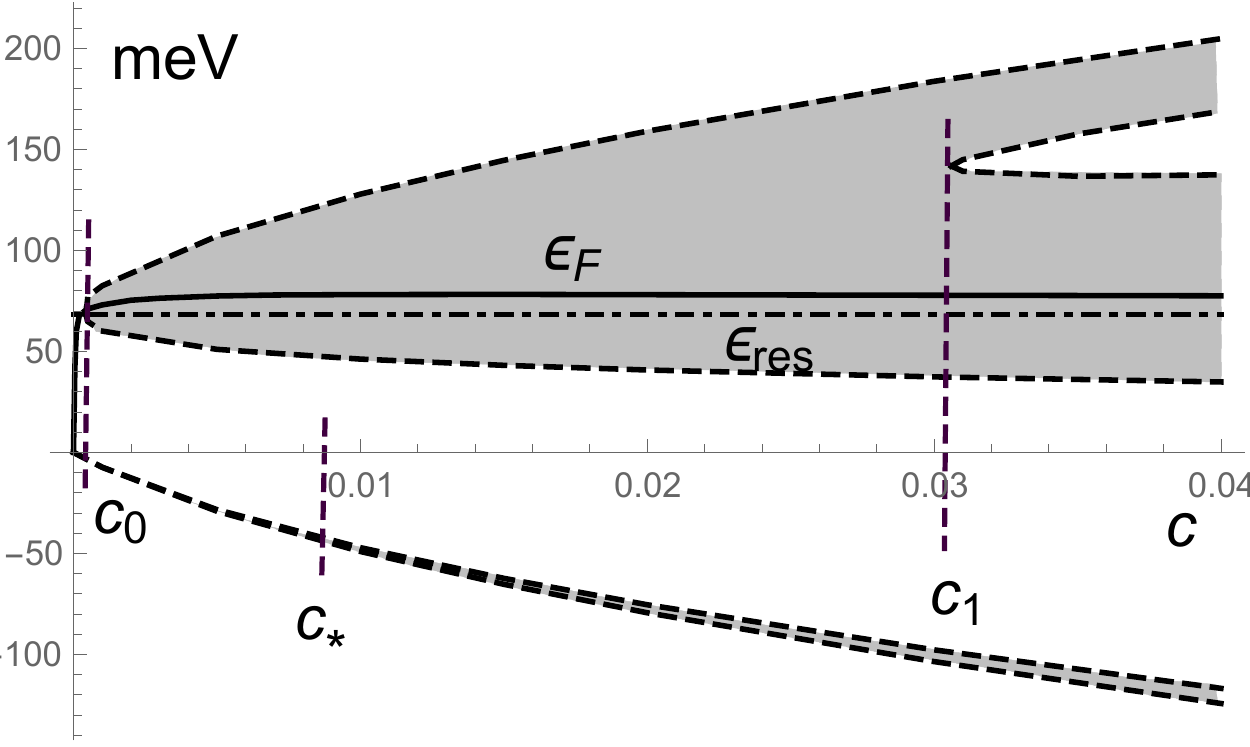}%Fig7
	\caption{Variation of mobility gaps (shadowed areas bordered by dashed lines) and Fermi level $\e_{\rm F}$ (solid line)
	{\it vs} concentration $c$ of Cu adatoms (the same as in Figs.~\ref{fig5}, \ref{fig6}). The mobility gap near $\e_{res}$
	onsets at $c_0 \approx 4\cdot 10^{-4}$, and widens until its $\D_{mob}$ and $\tilde\D_{mob}$ components being split at
	$c_1 \approx 3.1\cdot 10^{-2}$ by the emerging narrow impurity band. The Fermi level steeply grows from zero energy to
	enter 	$\D_{mob}$ (realizing a metal-insulator transition) just at $c = c_0$ and then stays close to $\e_{res}$.}
	\lb{fig7}
\end{figure}

With further growth of $c$, the IRM criterion can be continued using the complete (non-linearized) MPS given by
Eq.~\ref{qtop} in Eq.~\ref{IRMlim}. Multiple roots of the resulting equation are readily found numerically and the corresponding
mobility edges in function of $c$ are shown in Fig.~\ref{fig7}, for the same Cu t-impurities as in Figs.~\ref{fig5} and \ref{fig6}.
In particular, the critical concentration value following from Eq.~\ref{c0} for this case: $c_0 \approx 4\cdot10^{-4}$,
is well reproduced here. Also this picture shows how a sub-linear in $c$ growth of the composite mobility gap $\D_{mob} +
\tilde\D_{mob}$ gets eventually surpassed by a faster linear expansion of the impurity band, $E_{2,q}$, permitting its central
part to emerge from the localized range at the next critical concentration $c_1 \sim (\e_{res}/\tilde\o)^2 \gg c_0$. Physically,
this means the onset of a ballistic conductivity range in the spectrum from the insulating background.

Finally, a similar consideration holds for the vicinity of shifted Dirac point $\e_{\rm D}$, using the LMPS $q(\e)
\approx q_{max}|\e - \e_{\rm D}|/W$ in Eq.~\ref{IRMlim}, shows persistence of a very narrow mobility gap $\D_{\rm D} \approx
\hbar\t^{-1}(\e_{\rm D})$, even in the limit of $c \to 0$. This is due to vanishing l.h.s.~of Eq.~\ref{IRMlim}~here since
$q(\e_{\rm D}) = 0$, unlike that near $\tilde\e_{res}$ where $q(\tilde\e_{res})$ does not vanish even in the limit of $c \to
0$ and assures the IRM protection in this limit. The related gap grows as $\D_{\rm D} \sim (c^2c_0/c_\ast^3)\e_{res}$ until $c \ll
c_\ast = c_0/c_1$, then slowing down to $\D_{\rm D} \sim (c^{1/2}c_0/c_\ast^{3/2})\e_{res}$ at $c \gg c_\ast$, again in
a good agreement with the numerical result.

The general picture in Fig.~\ref{fig7} is yet properly completed with the plot of Fermi energy {\it vs} $c$ (obtained
by numerical integration of Eq.~\ref{total-dos} in Eq.~\ref{eF}). This process begins from its very fast advance as
$\e_{\rm F}(c) \approx \sqrt c W$ (resulting from integration of almost unperturbed DOS), from the initial $\e_{\rm F}(0)
= 0$ up to $\e_{res}$ vicinity, where this advance is abruptly hampered by the weight absorption into the resonance DOS peak.
After crossing the resonance level just at $c \approx c_0$ and entering the already formed mobility gap, the following very
slow $\e_{\rm F}(c)$ growth leaves it within the localized area (though it could be moved out of this narrow area, e.g., by
an electric bias). The resulting intermittency of localized and mobile states (metal-insulator and insulator-metal transitions)
within a narrow energy range around $\e_{res}$ can be of interest for applications.

At high enough concentrations, $c \gg c_0$, the resonance maximum of host DOS due to localized states near $\e_{res}$
is estimated as:
\bea
\r_{host}(\e_{res}) & \approx & \r_0(\e_{res})\left(1 + \frac 2\pi \arctan \frac c{c_0}\right.\nn\\
& + & \left. \frac c{\pi c_0}\ln\frac {\pi c_0}{c^2 + c_0^2}\right) \gg \r_0(\e_{res}),
\lb{rhost}
\eea
which is well pronounced against the linear graphene DOS, Eq.~\ref{re}, at this energy. This result also permits to
compare the spectral weights in the resonance range that stem from perturbed graphene host, $w_{host}$, and from AM
impurities themselves, $w_{imp}$. The integral weight of the resonance peak in $\r_{host}$ can be estimated as a
product of the resonance width $\G(\e_{res}) \approx \tilde\o^2\e_{res}/W^2$ and its height by Eq.~\ref{rhost},
giving $w_{host} \sim c \tilde\o^2/W^2\ln(1/c_0) \ll c$. The complementary weight, $w_{imp}$, can be approximated as
\bea
w_{imp} & = & \int_{\e_D}^{\tilde{\e}_{res}}\r_{imp}(\e)d\e \approx \int_{-\infty}^{\infty}\r_{imp}(\e)d\e\nn\\
& \approx & c\left(1 - \frac{\tilde\o^2}{W^2}\ln\frac1{c_0}\right).
\lb{wimp}
\eea
This shows that weakly coupled adatoms retain the main part of their total spectral weight $c$, having transferred only a
small rate to the delocalized bands. The dominant $\r_{imp}$ contribution to the total DOS $\r$ just within the localized
ranges is clearly seen in Fig.~\ref{fig6} (red curve).

At yet higher impurity concentrations, $c \gtrsim c_1$, the quasi-gap growth, though getting slower: $\tilde\e_{res}
- \e_{res} \approx \e_{res}(\sqrt{1 + 4c/c_1} -1)/2$, still stays faster of that for the mobility gaps, $\D_{mob} +
\tilde\D_{mob}$, keeping the same topology of mobility ranges in the low energy spectrum.

At least, the above employed T-matrix approximation for self-energy can be next justified by a more detailed treatment
of the non-trivial GE terms from Eq.~\ref{ge} (see Appendix \ref{A}) showing this approximation to stay sufficient down
to the established mobility limits. So the same MPS approach to the IRM criterion is extended for all the following impurity
types.

\subsection{Strongly coupled AM impurities, numerical studies}

It is of eminent interest to compare the above weak coupling AM results with the opposite limit of strong coupling.
First of all, this moves the impurity resonance $\e_{res}$ much closer to the initial Dirac point than the original
adatom on-site energy $\e_0$. Thus, for the example of H adatoms with strong $\o = 7.5$ eV coupling, their $\e_0
\approx 160$~meV gets reduced down to $\e_{res} \approx 6.9$~meV, see Fig.~\ref{figH}, compared to the Cu case with
$\o = 0.81$ eV, where $\e_0 \approx 80$~meV is only reduced to $\e_{res}\approx 69$~meV, seen in Fig.~\ref{fig6}.

Another striking difference between weakly and strongly coupled AM t-impurities is in the part of their total spectral
weight transferred to the electronic states of the host system. Comparing the red curves displaying $\rho_{imp}$,
Eq.~\ref{rimp}, in Figs.~\ref{fig6} and \ref{figH}, we see that weakly coupled Cu impurities retain larger spectral weight
around $\e_{res}$, while the strongly coupled H ones hold just a very tiny its fraction (in a slim peak centered at
$\e_{res}$).
\begin{figure}[h]
	\centering
	\includegraphics[scale=0.6]{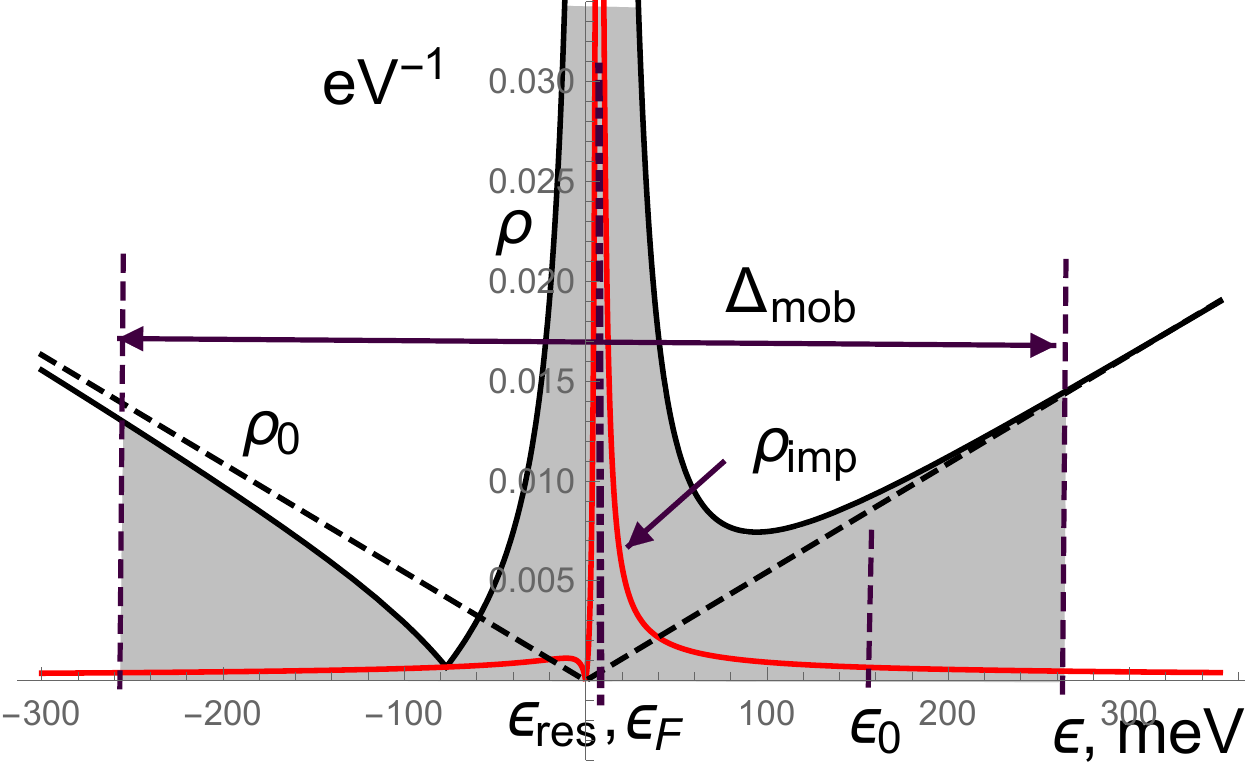}%Fig8
	\caption{Restructured DOS of graphene under H adatoms (see the AM parameters in Table~\ref{tab:table1}) with concentration
	$c = 0.01$ and symmetric sublattice occupation. The total value $\rho$ (black line) and its impurity component $\rho_{imp}$
	(red line) are compared to $\rho_0$ of unperturbed graphene (dashed line). The mobility gap $\D_{mob}$ range is shadowed.}
	\lb{figH}
\end{figure}
\begin{figure}[h]
	\centering
	\includegraphics[scale=0.6]{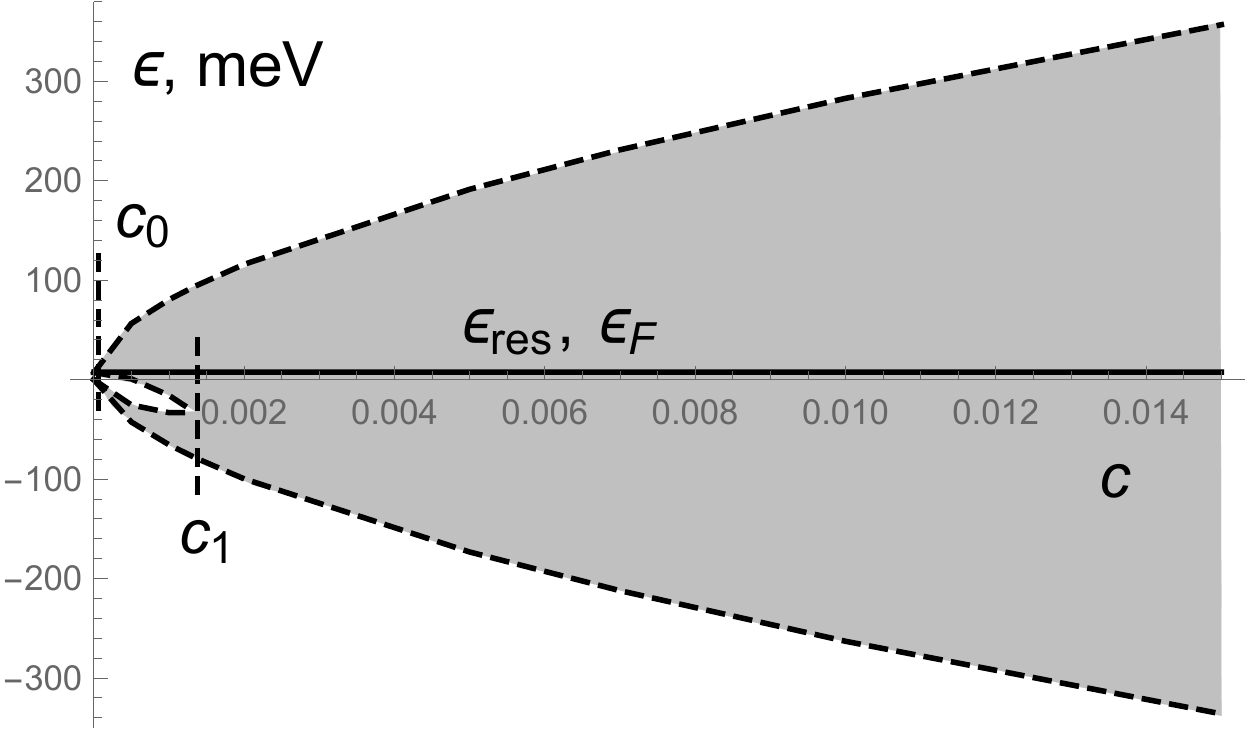}%Fig9
	\caption{Mobility gaps (shadowed areas bordered by dashed lines) and Fermi level (solid line) {\it vs} concentration
	$c$ of strongly coupled H adatoms, compare with the related DOS in Fig.~\ref{figH}. After the upper $\D_{mob}$ gap
	onsets and absorbs $\e_{\rm F}$ at extremely low $c_0 \approx 4\cdot 10^{-6}$, it rapidly merges with the lower
	$\D_{\rm D}$ already at $c_1 \approx 2\cdot 10^{-3}$.}
	\lb{fig7a}
\end{figure}

 Also a strong host-impurity coupling modifies the above estimates for the mobility gap near that resonance, making it much
 broader. Correspondingly, the Fermi level enters it at as low critical concentration as $c_0 \approx 4\cdot  10^{-6}$, for
 the H case, and then stays close to the resonance, as shown in Fig.~\ref{fig7a}. This makes the metallic state extremely
 unstable against such impurities (within the adopted graphene model with no intrinsic band splitting, e.g., by spin-orbit
 effects). At last, the strong impurity-host coupling favors to merging of different mobility gaps observed in the weak coupling
 case, as seen in a rapid absorption of the narrow $\D_{\rm D}$ by much broader $\D_{mob}$ in Fig.~\ref{fig7a} and no traces for
 decoupling of $\tilde\D_{mob}$.

Depending on the sign of the on-site energy $\e_0$, the resonance $\e_{res}$ develops below or above the graphene charge
neutrality (Dirac) point. For two considered AM cases, Cu and H, they lie above, and those situations resemble donor-like
dopants in common semiconductors---the total carrier weight determining the Fermi level, see Eq.~\ref{eF}, is $Q = 1 + c>1$.
For the case of F, negative $\e_0 = -2.2$~eV leads to $\e_{res} \approx -0.4$~eV, and the whole situation resembles acceptor-like
dopants, where the carrier weight turns $Q = 1 - c < 1$. This produces the DOS picture as displayed in Fig.~\ref{DOS_F}, seen
qualitatively as a mirror to the cases of donor impurities ($\e_0 > 0$), and so the restructured spectrum is of inverted type.

\begin{figure}[h]
	\centering
	\includegraphics[scale=0.6]{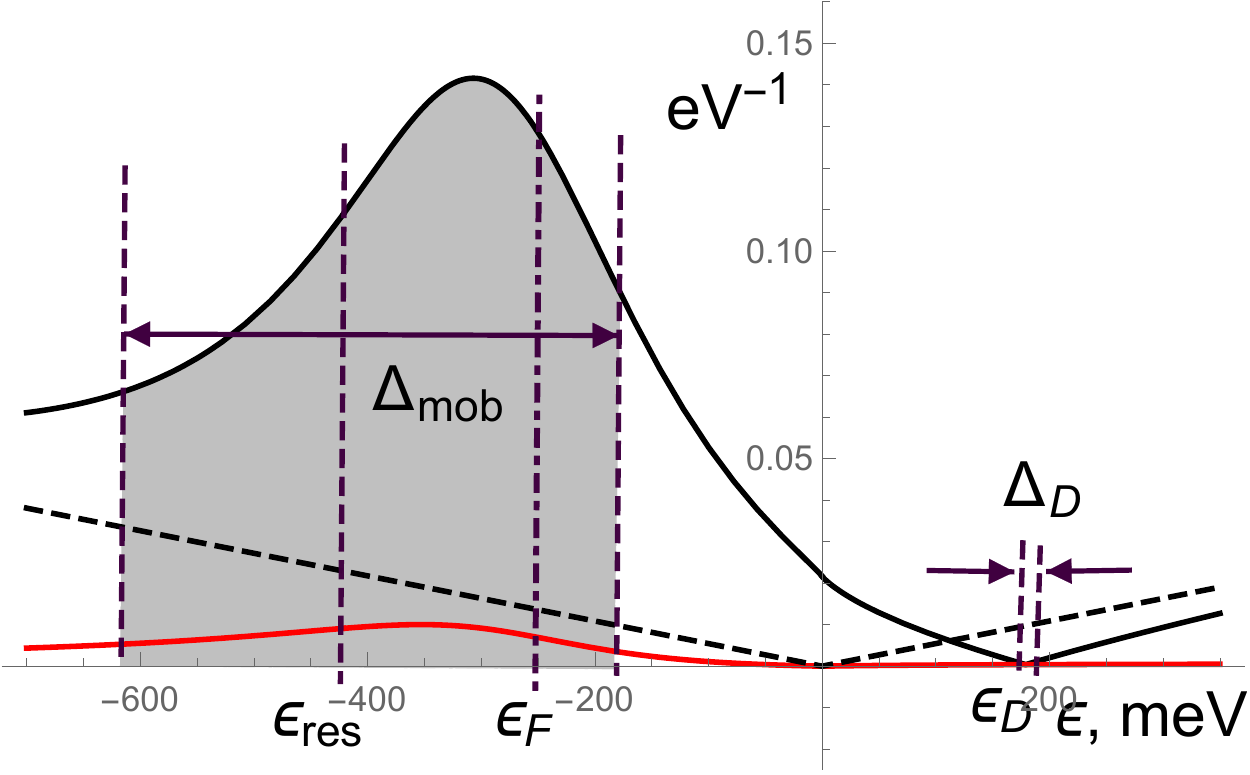}%Fig10
	\caption{Restructured DOS of graphene under F adatoms with concentration $c = 0.03$ and symmetric sublattice occupation
	(see the model parameters in Table~\ref{tab:table1}). The total value $\rho$ (solid line) is compared to the unperturbed
	graphene linear DOS, $\rho_0$ (dashed line), and two shaded areas present mobility gaps $\D_{mob}$, and $\D_{\rm D}$.}
	\lb{DOS_F}
\end{figure}

\begin{figure}[h]
	\centering
	\includegraphics[scale=0.6]{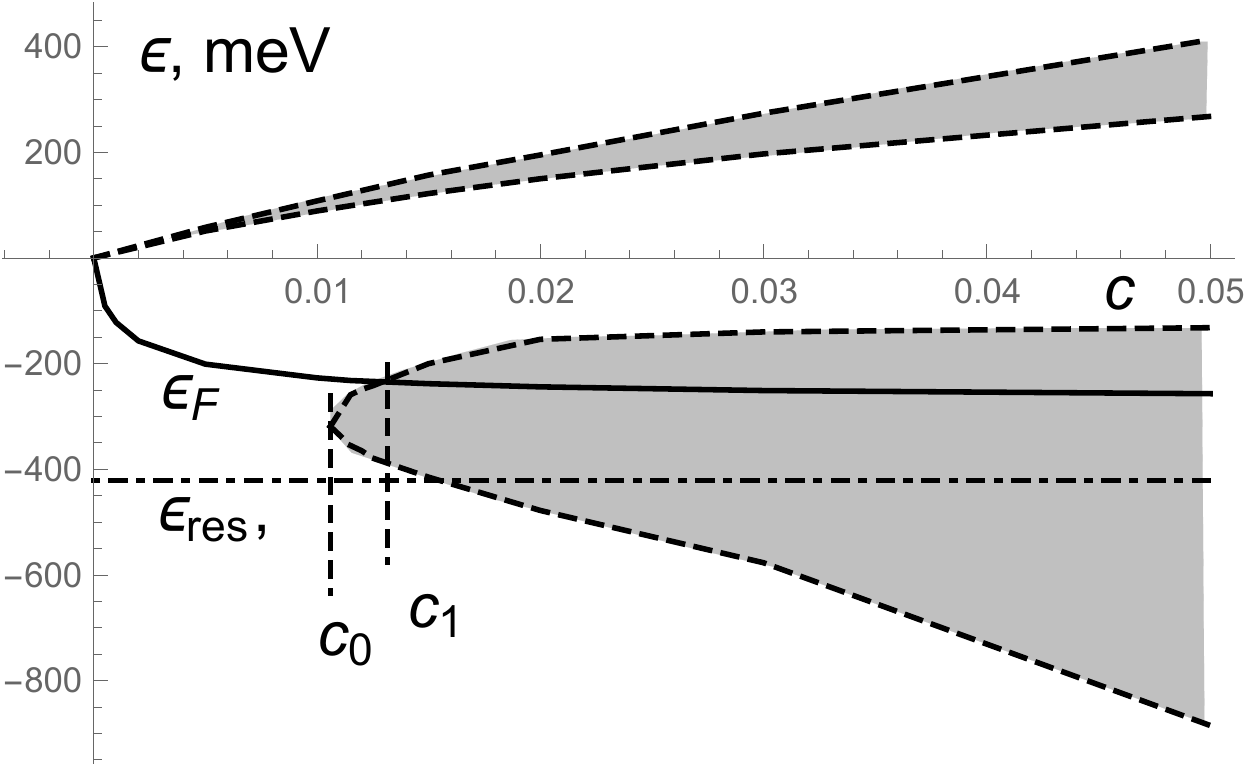}%Fig11
	\caption{Mobility gaps, bottom $\D_{mob}$ and top $\D_{\rm D}$ (shadowed areas bordered by dashed lines) and the Fermi level
	(solid line) {\it vs} concentration $c$ of F adatoms in graphene. The localized range $\D_{mob}$ onsets near the resonance
	energy $\e_{res} \approx -0.4$~eV at the critical concentration $c_0 \approx 1.06\cdot 10^{-2}$ to absorb the Fermi level at
	the next critical value  $c_1 \approx 1.3\cdot 10^{-2}$, realizing a robust metal-insulator transition for hole-type charge
	carriers.}
	\lb{Mob_F}
\end{figure}

Here, with growing the impurity concentration $c$, the Fermi level goes monotonously {\it down} from zero and enters the mobility
gap near the impurity resonance at some $c_1 > c_0$, which results in a robust metal-insulator transition for the hole-type charge
carriers, see Fig.~\ref{Mob_F}. Those results are in agreement with the experimental findings of Hong \emph{et~al}~\cite{Hong2011}
that report metal-insulator transition in the fluorinated graphene at certain charge doping levels.

\subsection{Asymmetric t-occupancy}

Alike that for non-resonant LM impurities, the above discussed effects for AM t-impurities get altered when considering
asymmetric sublattice occupations. In this section we only focus on the extreme case corresponding to $c_1 = c$, $c_2 = 0$.
\begin{figure}[h!]
	\centering
	\includegraphics[scale=0.6]{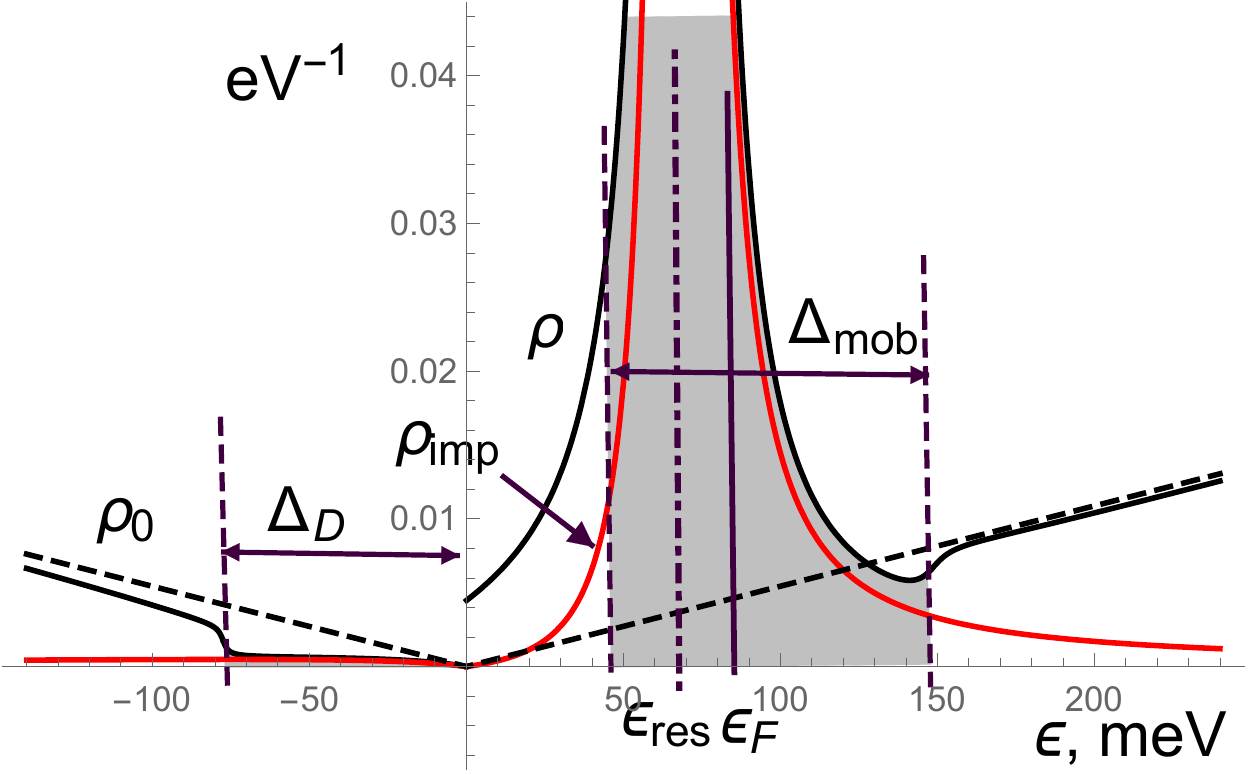}%Fig12
	\caption{Density of states for graphene with Cu t-impurities at their concentration $c = 0.01$ and asymmetric
	occupation 	of host sublattices (the same notations used for its elements as in Fig.~\ref{fig6}).
	%\textcolor{red}{Yuriy, what is the concentration here? The same $c=0.01$?}
	}
	\lb{tas}
\end{figure}

\begin{figure}[h!]
	\centering
	\includegraphics[scale=0.6]{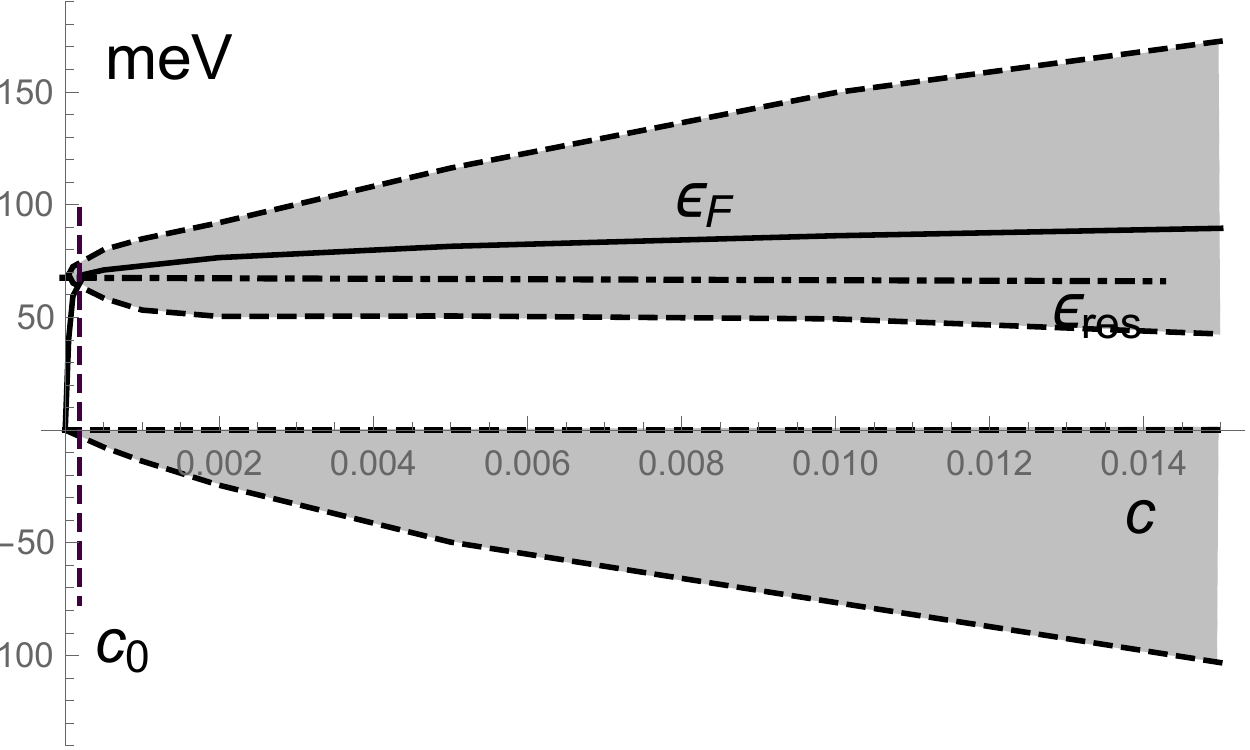}%Fig13
	\caption{Mobility gaps (shadowed areas bordered by dashed lines) and Fermi level (solid line) {\it vs} concentration
	$c$ for Cu adatoms at asymmetric occupation.
	Note the difference of mobility gaps development from the case of symmetric occupation in Fig.~\ref{fig7}.}
	\lb{Mobas}
\end{figure}

Having the related T-matrix: $\hat T(\e) = T_t(\e)\hat m_1$, the direct evaluation of ${\rm Re}\bigl[{\rm Det}\,
\hat G_\bk^{-1}(\e)\bigr]$ with $\hat G_\bk^{-1}(\e)$ given by Eq.~\ref{seAMtop} results in the following secular
equation:
\be
\e^2 - \e_q^2 - 2c\,\e\, {\rm Re\,}\,T_t(\e) = 0,
\lb{das}
\ee
that gives the restructured energy spectrum. Linearizing $T_t(\e)$ in the above expression turns it into the cubic
equation with respect to $\e$, unlike the symmetric t-case governed by Eq.~\ref{chaAMtop1}. The EPS roots $E_{\a,q}$
($\a = 1,2,3$) of Eq.~\ref{das} can be straightforwardly obtained by the Cardano's formulas, but their following analyzes
turn to be awkward and unpractical. However, the above secular equation also admits an easy and ``user-friendly'' MPS:
\be
q(\e) = \frac{q_{max}}W\sqrt{\e^2 - 2c\,\e\,T_t(\e)},
\lb{dispas}
\ee
which leads to the corresponding DOS:
\bea
\r_{as}(\e) & = &\frac2{\pi W^2}{\rm Im\,}\left\{\left[\e - c\e T_t(\e)\right]\right.\nn\\
&\times& \left.\ln\left[1 - \frac {W^2}{\e^2 - 2c\,\e\,T_t(\e)}\right]\right\},
\lb{dosas}
\eea
presented in Fig.~\ref{tas}. Its main difference from the symmetric counterpart, Fig.~\ref{fig6}, consists in the
opening of an effective gap, $\D_{\rm D}$, from the initial zero Dirac point to its shifted position, $\e_{\rm D} \approx
-2c\tilde\o^2/\e_{res}$, alike the case of asymmetric occupancy in LM, displayed in Fig.~\ref{fig3}. Next, using the MPS
by Eq.~\ref{dispas} in the IRM criterion by Eq.~\ref{IRMlim}, one can estimate the underlying mobility gaps, as well near the
subband edges as around the resonance peak. The corresponding subbands and mobility gaps are displayed in function of
impurity concentration in Fig.~\ref{Mobas}. From the point of view of metal-insulator transitions, the asymmetric AM
scenario offers a richer intermittency between the extended and localized ranges and, along with presence of a wide and
almost pure $\D_{\rm D}$ gap in its spectrum, it is expected to provide a more promising application-oriented playground
than the symmetric case.

\vspace{3 mm}

\section{Anderson's impurities at bridge and hollow positions}
\lb{pos2}

\subsection{\label{sec:bridge}Bridge position}
Practically the same scenario as for symmetric AM t-impurities is found for AM impurities at b-positions, though this
conclusion requires some additional analysis and clarification.

Assume an AM impurity to occupy a b-position projected at $\br$, then its two neighboring carbon atoms reside at host
sites $\bn_{1,i}  = \br - \bde_i/2$ (A sublattice) and $\bn_{2,i}=\br+\bde_i/2$ (B sublattice), where $\bde_i$ is one
of three nearest neighbor vectors defining the given bridge, see Figs.~\ref{fig1}~and~\ref{fig4}. The corresponding
scattering spinor in the conduction-valence band space, Eq.~\ref{ss}, is explicitly given as:
\be
u_{\br,\bk} = \sqrt 2\,{\rm e}^{i\bk\cdot\br}\,
\left(
\begin{array}{r}
  \cos{\tfrac{1}{2}(\bk\cdot\bde_i-\mathrm{arg}\,\g_{\bk})}\\
-i\sin{\tfrac{1}{2}(\bk\cdot\bde_i-\mathrm{arg}\,\g_{\bk})}
\end{array}
\right).
\lb{ssb}
\ee
Here $\bk$ is referred to the $\G$-point and the hopping factor argument, $\mathrm{arg}\,\g_{\bk}$, is given by Eq.~\ref{phi}.
This spinor defines the scattering matrix $\hat V_{\br,\bk,\bk'}$, Eq.~\ref{vef}, and then the momentum diagonal T-matrix,
Eq.~\ref{ttk}, as:
\bea
\hat{T}_{\bde_i,\bk}(\e) & = & T_t(\e)
\left[
\hat 1 + \hat \s_3 \cos{\left(\bk\cdot\bde_i - {\rm arg}\,\g_{\bk}\right)}
\right.
 \nn\\
& - &
\left.
\hat \s_2 \sin{\left(\bk\cdot\bde_i - {\rm arg}\,\g_{\bk}\right)}\right],
\lb{tmb}
\eea
with the same scalar prefactor $T_t(\e)$ as in the t-case, Eq.~\ref{tm}. Assuming also equal average occupancy of three
non-equivalent bridge configurations, $c_{\bde_i} = c/3$, the partial T-matrices $\hat T_{\bde_i}$ combine into the total
T-matrix:
\be
\hat T(\e,\bk) = \frac13\sum_{i=1}^3 \hat T_{\bde_i,\bk}(\e) = \left(\hat 1 + \frac{|\g_{\bk}|}{3}\,\hat \s_3\right)
T_t(\e).
\lb{tbk1}
\ee
For momenta $\bk$ close to the graphene valleys centers, $\bk = \bq + \bK^{(\prime)}$, one can employ the low-energy
expansion to present the T-matrix for b-impurities as:
\be
\hat T(\e,q) = \left(\hat 1 + \frac{\e_q}{3t}\,\hat \s_3\right)T_t(\e),
\lb{tbk}
\ee
thus dependent on the radial component $q$ of reduced momentum. But in the long-wave limit, $\e_q \ll W \sim 3t$, the
momentum dependent term in Eq.~\ref{tbk} can be practically neglected. Therefore in the considered low-energy limit, the
b-case T-matrix gets effectively reduced just to $T_t(\e) \hat 1$. As a consequence, the restructured energy spectrum
in the presence of b-impurities should mostly reproduce the same spectral features as for the symmetric t-case.

To what types of adatoms on graphene one can apply the above findings? First-principle calculations predict oxygen and
nitrogen to bond in the bridge position~\cite{Wu:APL2008}. However, also for some top positioned impurities, like copper~\cite{Wu:APL2009,Amft:JPhysCondMat2011,Frank:PRB2017}, and gold~\cite{Chan2008:PRB,Amft:JPhysCondMat2011} the
energy difference between the top and bridge configurations is relatively small, and therefore their bridge realization
can become probable. Similarly, the light ad-molecules like CO, NO and NO$_2$ prefer to adsorb~\cite{Leenaerts:PRB2008}
equally-likely to the hollow and bridge positions.

\subsection{\label{sec:hollow}Hollow position}

Hollow-type AM impurities represent a special case; an adatom in the h-position displays local $C_{6v}$ symmetry, which
strongly reduces the coupling of impurity degrees of freedom with host graphene states (see Eq.~\ref{tmh} below).
That was earlier interpreted as their full decoupling \cite{Ruiz2016} from graphene states. However, it will be shown below
that, when treated consistently within the AM, the h-impurities are sufficient to produce essential
restructuring of graphene low-energy spectrum. The resulting
h-type resonances and the related spectral features in terms of AM parameters are compared in what follows with the previously
discussed  t-~and~b-cases.

For an h-impurity projected to $\br$, the sum in the scattering spinor $u_{\br,\bk}$, Eq.~\ref{ss}, counts its 6 carbon
neighbors. Those are residing at host sites: $\bn_{1,i} = \br + \bde_i$ (A sublattice), and $\bn_{2,i} = \br-\bde_i$ (B
sublattice, see Figs.~\ref{fig1}~and~\ref{fig4}). This summation results in:
\be
u_{\br,\bk} = \sqrt{2}\,|\g_{\bk}|\,{\rm  e}^{i\bk\cdot \br} \left(
\begin{array}{r}
 \cos{\bigl(\tfrac{3}{2}\,\mathrm{arg}\,\g_{\bk}\bigr)}\\
i\sin{\bigl(\tfrac{3}{2}\,\mathrm{arg}\,\g_{\bk}\bigr)}
\end{array}\right).
\lb{ssb}
\ee
Implementing this into Eq.~\ref{ttk} leads to the corresponding momentum diagonal T-matrix:
\bea
\lb{tmh}
\hat T_{\br,\bk}(\e) & = & T_h(\e)\,|\g_{\bk}|^2\,\Bigl[\hat 1 + \hat \s_3 \cos{\left(3\,\mathrm{arg}\,\g_{\bk}\right)}
\nn\\
& & \qquad + \hat \s_2 \sin{\left(3\,\mathrm{arg}\,\g_{\bk}\right)} \Bigr].
\eea
where the scalar prefactor $T_h(\e) = \o^2/D(\e)$ differs from $T_t(\e)$, Eq.~\ref{tm}, by more complex denominator:
\be
D(\e) = \e - \e_0 + \frac{2\e\o^2}{t^2}\bigl[1 - \e G^{(0)}(\e)\bigr].
\lb{dh}
\ee
Similarly to the b-case, Eq.~\ref{tbk}, the h-impurity T-matrix, Eq.~\ref{tmh}, depends apart of the radial momentum $q$,
also on its azimuthal component encoded in $\mathrm{arg}\,\g_{\bk}$. This makes the restructured dispersion relation based
on Eq.~\ref{tmh} anisotropic. Another important difference of the h-case T-matrix from the t- and b-cases is in the small
prefactor, $|\g_{\bq+\bK^{(\prime)}}|^2 \approx (\e_q/t)^2\ll 1$, in its numerator, which is responsible for the above mentioned
decoupling of the graphene low-energy states with h-type AM impurities. The complete low-energy T-matrix for momenta near
the $\bK^{(\prime)}$ point reads:
\be
\hat T(\e,\bq) =T_h(\e)\frac{\e_q^2}{t^2}\Bigl[\hat 1 \mp \hat \s_3 \cos{\left(3\theta_\bq\right)} - \hat \s_2
\sin{\left(3\theta_\bq\right)} \Bigr],
\lb{tmh1}
\ee
where the plus (minus) sign applies to $\bK^{(\prime)}$ valley, and the form of angle $\theta_\bq$ is given by Eq.~\ref{phi}.

\begin{figure*}
\centering
\includegraphics[width=0.99\textwidth]{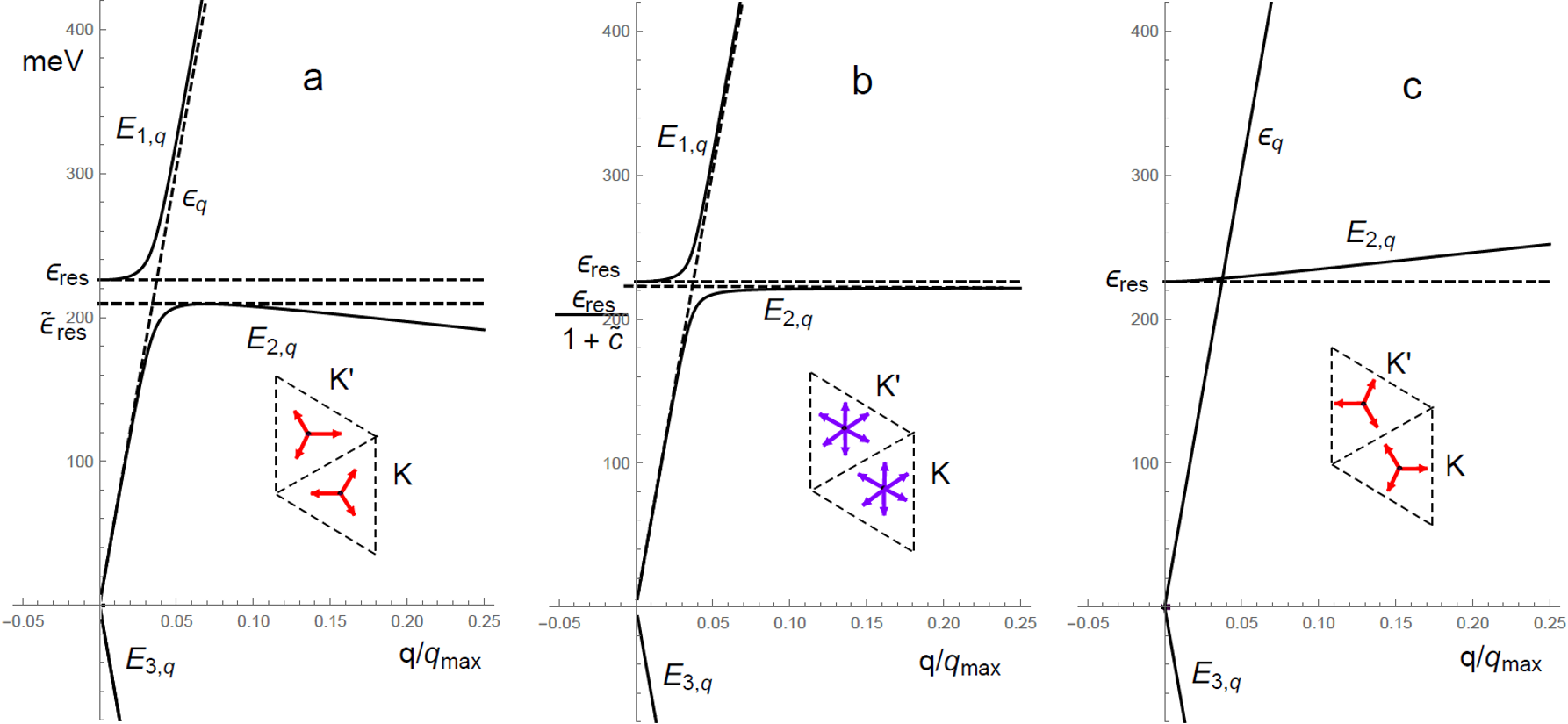}%Fig14
\caption{Restructured electronic dispersion of graphene in the presence of h-positioned impurities with $\e_{res} \approx 226$ meV, $\o = t/\sqrt 2$, and
concentration $c = 0.02$, along the nodal (blue arrows) and anti-nodal (red arrows) directions in the BZ with respect to $\bK$ and $\bK^{(\prime)}$ points, see insets.}
\label{figh123}
\end{figure*}

The general formulas, Eqs.~\ref{tmh}-\ref{tmh1}, allow to study, at least numerically, the spectral effects of h-type AM impurities
in a broad energy range. However, in what follows we stay rather on analytical side, using proper approximations near the Dirac
points. For example, to find the resonance pole of T-matrix and the restructured dispersion laws over the low-energy range,
$(\e/t)^2 \ll 1$, it is well justified to ignore the strongly suppressed $G^{(0)}(\e)$ term in the denominator $D(\e)$, which can be
then approximated by $D(\e)\approx \e - \e_0 + 2\e\o^2/t^2$. The correspondingly
approximated $T_h(\e) \approx \tilde\o^2/(\e - \e_{res})$ involves the resonance level:
\be
\e_{res} \approx \frac{\e_0}{1 + 2(\o/t)^2},
\lb{resh}
\ee
and the effective coupling constant $\tilde\o^2 = \o^2t^2/(t^2 + 2\o^2)$. Then the secular equation, Eq.~\ref{chaAMtopRe}, takes the form of
an ordinary cubic equation:
\be
\e^2 - \e_q^2\left(1 + \tilde c \,\frac{\e - \e_q \cos 3\theta_\bq}{\e - \e_{res}}\right) = 0,
\lb{disph}
\ee
with $\tilde c = 2c(\tilde\o/t)^2$. As noted above, the resulting dispersion is anisotropic, and the angular $\cos 3\theta_\bq$
dependence imprints the spectrum near $\e_{res}$ the $C_3$ symmetry. In sequel we characterize that general spectrum by its behavior
along the basic directions in the momentum plane: the nodal with $\cos3\theta_\bq = 0$, and the anti-nodal
with $\cos3\theta_\bq = \pm 1$. The main features for each considered case are shown in Fig.~\ref{figh123} and can be
summarized as follows.

i) Along the anti-nodal directions: $\theta_\bq = \pi,\pm \pi/3$ around the $\bK$-point and $\theta_\bq = 0,\pm 2\pi/3$
around the $\bK'$-point (red arrows in Fig.~\ref{figh123}a), the restructured spectra:
\bea
E_{^1_2,\bq} & = & \frac{\e_{res} + \e_q \pm \sqrt{(\e_{res} - \e_q)^2 + 4 \tilde{c}\e_q^2}}{2},\nn\\
E_{3,\bq} & = & -\e_q,
\lb{disph123}
\eea
include the purely unperturbed valence graphene band $-\e_q$, and the restructured $E_{^1_2,q}$ bands. They emerge from the spectral repulsion
between the graphene conduction band $\e_q$ and the resonance level $\e_{res}$ (supposing for definiteness $\e_{res} > 0$).
The most notable features here are the formation of a wider quasi-gap (anti-crossing) between $\e_{res}$ and $\tilde\e_{res} = \e_{res}/(1 + 4\tilde{c})$
and the inverted group velocity of $E_{2,q}$ at $\e_q > 2\tilde\e_{res}$.
This is due to the $q^2$-growth of the effective impurity-host coupling. Inverted group velocity generates also lower impurity side-band $W_{imp} \approx \tilde{c}W$, see Figs.~\ref{fig9} and~\ref{Hol2}, which is still broad enough compared to the related mobility gap $\Delta_{mob}$.

ii) Along the nodal directions: $\theta_\bq = \pm\pi/6$, $\pm\pi/2$, $\pm 5\pi/6$ (shown by blue arrows around each
$\bK$-point in Fig.~\ref{figh123}b), the cubic equation, Eq.~\ref{disph}, promotes couplings of the resonance level $\e_{res}$ to
both graphene $\pm\e_\bq$ bands. On one side the strong interaction of $\e_{res}$ with the conduction band $\e_q$ produces two
restructured bands, $E_{^1_2,q}$, with a very narrow anti-crossing between their asymptotic limits $E_{1,0} = \e_{res}$ and $E_{2,q_{max}} = \e_{res}/(1 + \tilde c)$.
Contrary, a weak non-resonant coupling of $\e_{res}$ with the valence band $-\e_q$ results only in slight modification of the latter, band $E_{3,q}$.

iii) Along the inverted anti-nodal directions: $\theta_\bq = 0,\pm 2\pi/3$ around the $\bK$-point and $\theta_\bq = \pi,\pm
\pi/3$ around the $\bK'$-point (red arrows in Fig.~\ref{figh123}c), these spectra:
\bea
E_{1,\bq}&=&\e_q,\nn\\
E_{^2_3,\bq}&=&\frac{\e_{res} - \e_q \pm \sqrt{(\e_{res} + \e_q)^2 + 4\tilde{c}\e_q^2}}{2},
\lb{disph123}
\eea
include the purely unperturbed graphene conduction band $\e_q$, and subbands $E_{2,q}$ and $E_{3,q}$ that originate from
a non-resonant repulsion between $-\e_q$ and $\e_{res}$. Subband $E_{2,q}$ has the width $W_{imp}'$, see Figs.~\ref{fig9}
and~\ref{Hol2}, and the valence $E_{3,q}$ only slightly deviates from the original valence band $-\e_q$.

Another peculiarity here is the absence of the shift of the energy level for the Dirac point that was present as $\e_{\rm D}
\neq 0$ in the previous cases. Also peculiar DOS features appear near the impurity resonance, as shown in Fig.~\ref{fig9},
with their notable differences from the t- and b-cases. First, in practical vanishing of quasi-gap (due to the same small
prefactor in the impurity-host coupling as indicated before Eq. \ref{tmh1}) and second, in the appearance of new side-bands
around $\e_{res}$ with widths $\approx \tilde c W$, that can be seen as the ``impurity induced heavy fermions'' with an
emergent $f$-wave symmetry. The details of analytic calculation of this DOS function are given in Appendix \ref{B}.

The resulting sum of the DOS components, $\r_{host}(\e) = \pi^{-1}{\rm Im}\,g(\e)$ and $\r_{imp} = {\rm Im}\,
T_h(\e)$ presented in Fig.~\ref{fig9} reveals its contributions from the spectrum branches $E_{^1_2,\bq}$ with a spike at
$\e_{res}$ and a break at $\tilde\e_{res}$.
\begin{figure}[h!]
	\centering
	\includegraphics[scale=0.65]{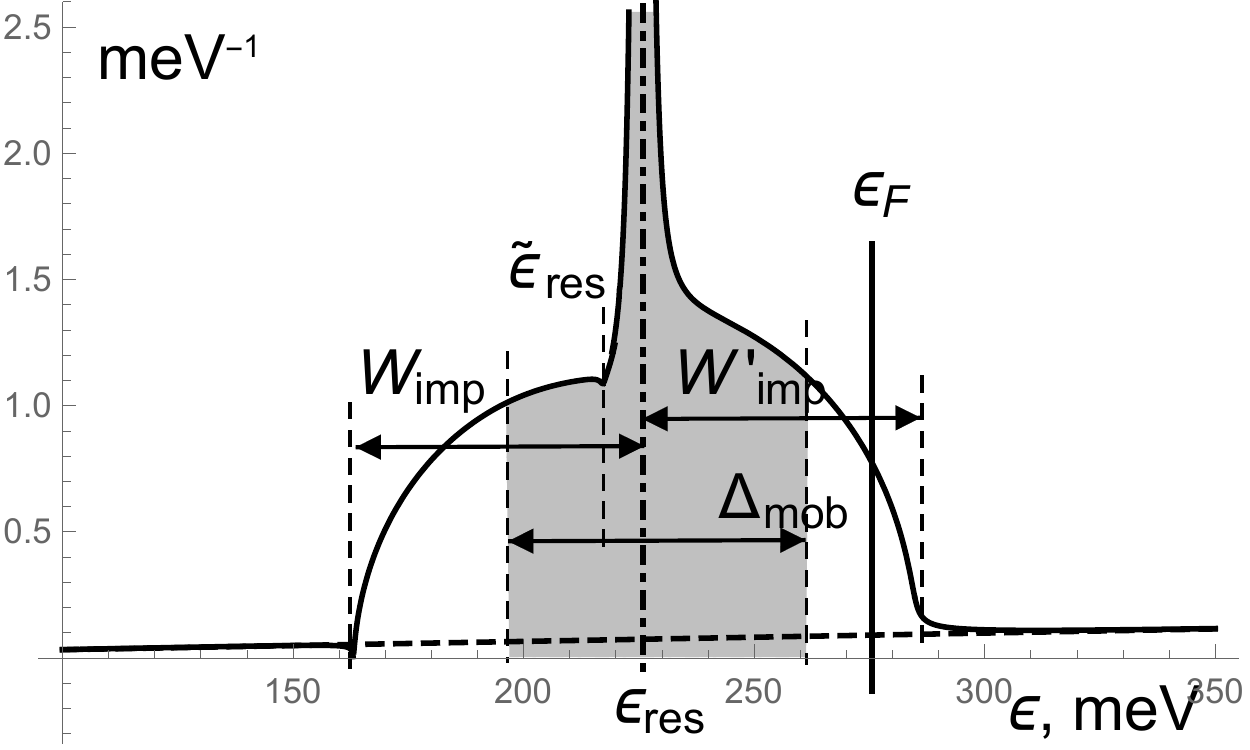}%Fig15
	\caption{DOS due to h-position impurities with AM parameters as in Fig.~\ref{figh123}. The resonance peak at $\e_{res}$
	is bordered from both sides by much wider impurity side-bands $W_{imp}$ and $W'_{imp}$. The localized states within
	the mobility gap $\Delta_{mob}$ are shadowed, the unperturbed DOS $\r_0(\e)$ is shown by the dashed line.}
	\lb{fig9}
\end{figure}

The obtained dispersions and DOS can be further used for the IRM criterion, Eq.~\ref{IRMlim}, and for comparing the Fermi level
and mobility edge positions. In this approach, the dispersion equation in its complete form:
\be
\e^2 - \e_q^2\left[1 + \frac{2c\o^2}{t^2}\frac{\e - \e_q\cos 3\theta_\bq}{D(\e)}\right] = 0
\lb{disph2}
\ee
(instead of simplified Eq.~\ref{disph}), can provide an MPS $q(\e,\theta)$ along a given azimuthal direction $\theta=\theta_\bq$. Then the
related mobility edges can be estimated numerically from an extension of Eq.~\ref{IRMlim}:
\be
{\rm min}_\theta \,\left|\frac{{\rm Re}\,q(\e,\theta)}{\partial{\rm Re}\,q(\e,\theta)/\partial\e}\right| = \hbar\t^{-1}
(\e),
\lb{a6}
\ee
where, from symmetry considerations, the minimum is sought along the above defined nodal and anti-nodal directions.
Their comparison, readily, indicates such minimum to be along the anti-nodal directions displayed in Fig.~\ref{figh123}a (with
$\cos 3\theta_\bq = -1$ and the widest quasi-gap). The corresponding explicit solution of Eq.~\ref{disph2} reads:
\be
q_0(\e) = q_{max}\frac{\sqrt{D(\e)\left(D(\e) + 4c\e\right)} - D(\e)}{2cW},
 \lb{a8}
\ee
and using it in Eq.~\ref{a6} gives finally the mobility edges as shown in Fig.~\ref{Hol2}.
\begin{figure}[h!]
	\centering
	\includegraphics[scale=0.65]{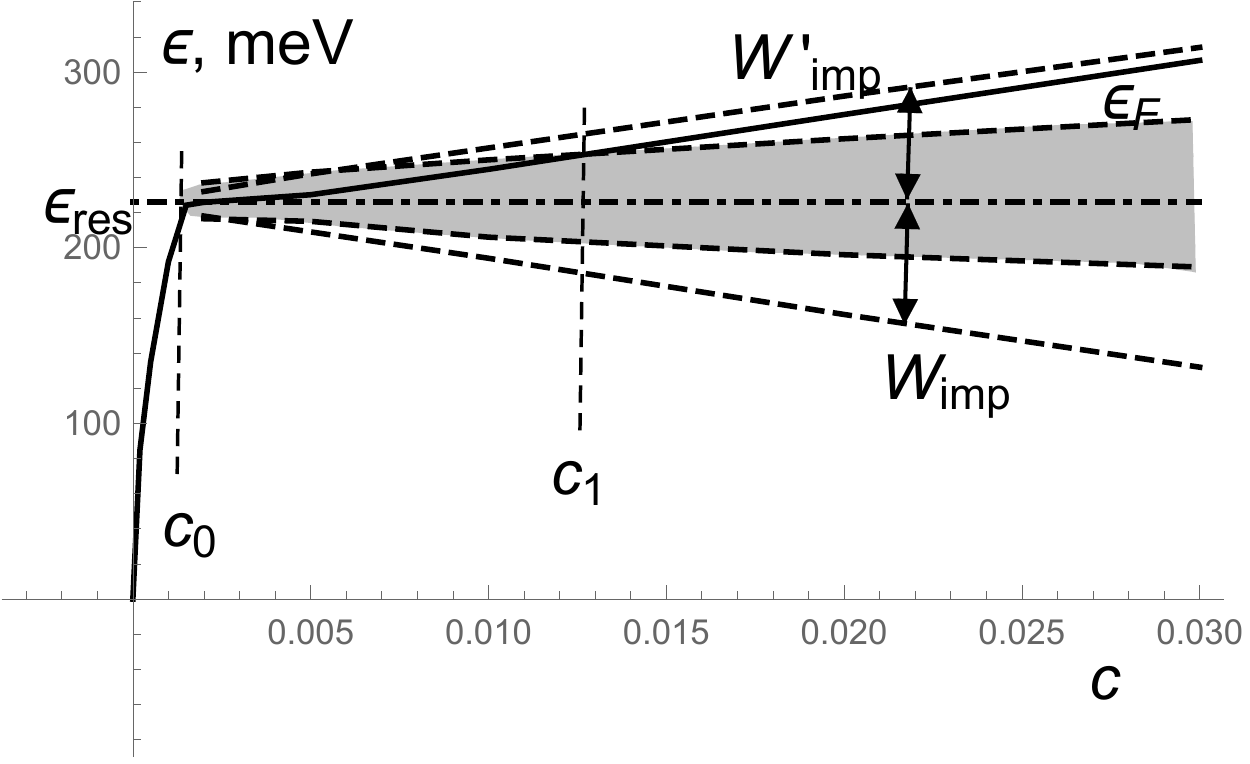}%Fig16
	\caption{Development of the mobility gap $\D_{mob}$, impurity side-bands $W_{imp}$ and $W'_{imp}$ and Fermi
	level $\e_{\rm F}$ {\it vs} concentration $c$ for h-position impurities with AM parameters as in
	Figs.~\ref{figh123}, \ref{fig9}.}
	\lb{Hol2}
\end{figure}
In similarity to the before considered t-cases displayed in Figs.~\ref{fig7}, \ref{fig7a}, \ref{Mob_F}, \ref{Mobas}, here
a localized range emerges near $\e_{res}$ at the critical concentration $c_0 \approx 1.5\cdot 10^{-3}$, and then extends
further sublinearly in $c$, see the shadowed area in Fig.~\ref{Hol2}. Its limits are exceeded from below and above by the
linearly growing outer side-bands $W_{imp}$ and $W'_{imp}$ (dashed lines) that contain extended ``heavy fermionic'' states,
see also the DOS displayed in Fig.~\ref{fig9}.
In this course, the Fermi level $\e_{\rm F}$ rises from zero through the initial conduction band $\e_q$
and then enters into the mobility gap at $c \approx c_0$. Further, with grown $c$ the Fermi level leaves that localized
region at another critical concentration $c_1 \approx 0.013$, and penetrates into the $W'_{imp}$ side-band with
the ``heavy fermionic'' character.
Thus h-type AM impurities realize both metal/insulator and insulator/metal transitions, but the two metallic phases
are different, the initial is $s$-like and the later is $f$-like.
Thus, the h-type adatoms can be considered, together with the asymmetric t-ones, as the most prospective candidates
for possible applications.

\textit{Ab-initio} studies are unveiling that light metallic adatoms~\cite{Chan2008:PRB} from groups I-III and
also heavy transition metals~\cite{Chan2008:PRB,Weeks2011:PRX,Mao:JOPCondMatt2008} are favored for adsorption above
the centers of graphene hexagons, i.e.~at hollow positions. The same is true for light ad-molecules like NH$_3$,
H$_2$O, NO$_2$ \cite{Leenaerts:PRB2008}.

\section{Discussion}
\lb{Disc}

The presented results demonstrate several characteristic tendencies that can accompany spectral transformation of the electronic band structure of graphene in the presence of disorder produced by impurities. The first decisive factor in that process is to
understand whether a single impurity center can produce a resonance energy level in the spectrum. The affirmative answer is
imposing some additional restrictions on the strengths of hybridization parameter and on-site energy. As demonstrated
above, this is practically always granted within the scope of Anderson hybrid model, reasonably justified for most of
common adatoms (ad-molecules) chemisorbed at graphene layer, and less granted for the isotopic Lifshitz model.

Once a resonance level exists for a given impurity type, by increasing their concentration the graphene spectrum would restructuralize following a particular scenario. The later is determined by the impurity locations [top, bridge, or hollow], and by their sublattice occupations [symmetric or asymmetric].
The most important spectral changes are emergencies of particular localized ranges and (pseudo)gaps that pop out inside the initial continuum of band states. Typically near the original resonance level, and also near the restructured Dirac points. Their further development with increased concentration is conveyed by splittings or mergings, as manifested by the fate of mobility edges that come from the phenomenological IRM criterion. It should be yet noted that such localized
ranges and related mobility gaps in the spectrum can also arise from a specific braking of the sublattice occupation of graphene
due to impurities, even, in the absence of single impurity resonance.
Due to its simplicity, the latter mechanism can be especially helpful in the search for practical realizations of properly restructured spectra.

The underlying electronic phase [metallic or insulating] of the resulting physical system is then essentially determined by the position of the Fermi level relative to the localized ranges. Those imprint the impurity type and concentration, but could be yet tuned by the external means, namely, electric or magnetic bias, temperature, etc., opening a wide field for possible applications.
Compared to the common situation in doped semiconductors, this provides much more versatile possibilities for interchange
of different types of metallic and insulating states, and mutual transitions among them. Also, in this course, there are
possibilities to combine the several spectral effects originating from different impurity species simultaneously, and thus target different energy ranges. However, the presented analysis did not consider the situation when randomly distributed impurities
at low concentration nucleate in nearest neighbor positions forming impurity clusters. Those in reality exist
(as known for some dopants in common semiconductors), and such direct impurity-impurity coupling will produce split resonances
and, correspondingly, more complicated series of localized energy ranges around them.

Finally, besides the purely electronic properties the variety of spectral regimes permits also other notable effects that employ additional degrees of freedom as, for instance, collective plasmonic spectra by narrow conduction bands, optical susceptibility by narrow
insulating gaps, Hall effect and magneto-transport on anisotropic Fermi surface, etc. From the above analysis the promising impurity types are weakly coupled t-position adatoms (including their donor-acceptor combinations)
and h-position hybridizing species (admitting a wider range of their atomic levels and coupling parameters).

The approach as presented, and the list of impurity effects that count the simplest host, single-layer graphene, can be further substantially developed in several different directions, for example, to multilayered graphene and its hexagonal lattice analogs, topological edge states, and quantum Hall effect regimes, Moir{\'e} patterns from plane rotations, etc. Such systems can present new playground for probing the interplay between the impurity disorder/localization effects, and the symmetry/topology order protection.

\section{Acknowledgements}
\lb{Ack}
The work of VML was partially supported by the Ukrainian-Israeli Scientific Research Program of the Ministry of Education
and Science of Ukraine and the Ministry of Science and Technology of the State of Israel, as well as by Grant Nos.~0117U000236
and 0117U000240 from the Department of Physics and Astronomy of the National Academy of Sciences of Ukraine. DK~acknowledges
support from Deutsche~Forschungsgemeinschaft, Project-ID~314695032 (SFB~1277), and the EU~Seventh~Framework~Programme under
Grant~Agreement~No.~604391~(Graphene~Flagship).

\vspace{1cm}
\bibliography{soc_ped}

\appendix
\section{Group expansion analysis}
\lb{A}
The above analysis was based on the impurity averaged GF's within the simplest T-matrix approximation. Generally, one
needs to check the higher order of GE (in powers of $c$) for the self-energy, Eq.~\ref{ge}, and their potential
impact on the formerly obtained results. Here the principal point is the convergence criterion for GE series, justifying
its approximation by the T-matrix term. In what follows we provide estimates for the first non-trivial pair term of GE,
giving a self-energy correction to the second order in $c$. We approximate the convergence criterion as:
\be
c|B_\bq| < 1.
\lb{gec}
\ee
It should be noted that, due to $j$-orthogonality of the scattering matrices $\hat{m}_j$, Eq.~\ref{ma}, such pair scatterings
contribute to the momentum diagonal GF only for t-impurities belonging to the same $j$-th sublattice. Therefore, the total
self-energy matrix for the momentum-diagonal GF results to be additive in the sublattice $j$-indices:
$$\hat G_\bq^{-1} =
\left(\hat G_\bq^{(0)}\right)^{-1} - \sum_j c_j\hat m_j\S_{j,\bq}.
$$
Each $j$-th sublattice self-energy $\S_{j,\bq}$ has
its own GE, analogous to general Eq.~\ref{ge}, with the corresponding pair term, $c_j B_\bq$, whose scalar B-factor is
explicitly given as follows \cite{Lifshitz}:
\be
B_{\bq}(\e) = \sum_{\bn \neq 0}\frac{{\rm e}^{-i\bq\cdot\bn}A_\bn(\e) + A_\bn(\e) A_{-\bn}(\e)}{1 - A_\bn(\e) A_{-\bn}
(\e)}.
\lb{pair}
\ee
This sum describes all multiple scatterings on impurity pairs from the same sublattice separated by lattice vectors
$\bn \neq 0$ (measured in units of graphene lattice constant $a$), returning a quasiparticle to its initial $\bq$-state,
through the dimensionless correlator:
\be
A_{\bn}(\e) = \frac{T_t(\e)}{N} \sum_{\bk} {\rm e}^{i\bk\cdot\bn}\,{\rm Tr}\,\hat G_{\bk}^{(0)}(\e).
\lb{are}
\ee
The later can be presented as a product:
$$
A_{\bn}(\e) = T_t(\e)\s_\bn f_\bn(\e),
$$
where the factor $\s_\bn = ({\rm e}^{i\bK \cdot\bn} + {\rm e}^{i\bK'\cdot\bn})/2$ takes the values 1, $z = {\rm e}^{2i\pi/3}$ and $z^\ast$ with the host lattice periodicity, as a consequence $\s_\bn \s_{-\bn} = 1$ (see Fig.~\ref{figzz1}).
\begin{figure}[h!]
	\centering
	\includegraphics[scale=0.35]{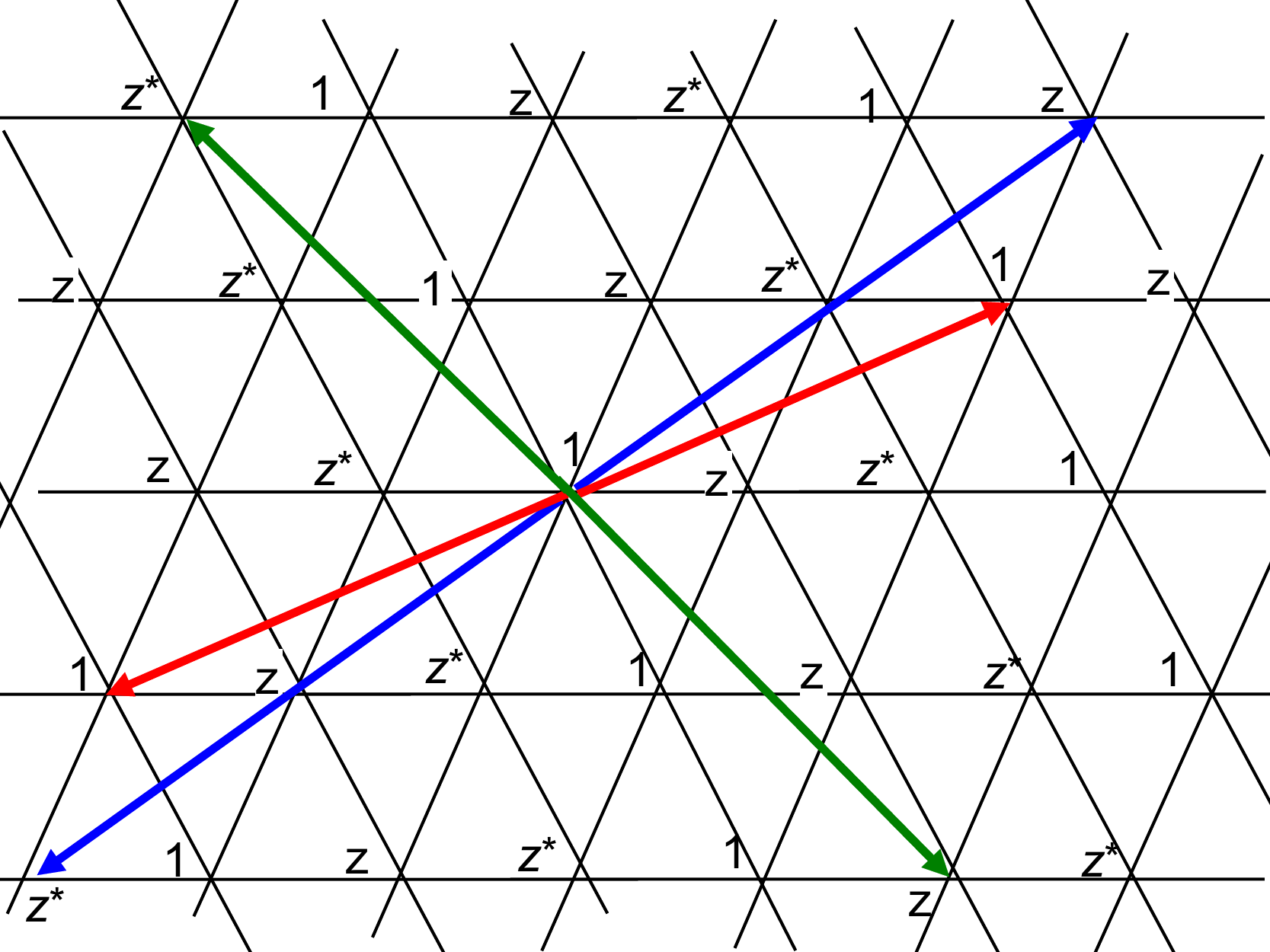}%Fig17
	\caption{Distribution of $\s_\bn$ values 1, $z = {\rm e}^{2i\pi/3}$ and $z^\ast$ over lattice sites (from the same
	$j$-sublattice) with examples of their products in opposite pairs (with respect to an initial zero site) to
	satisfy $\s_\bn \s_{-\bn} = 1$.}
	\lb{figzz1}
\end{figure}
The remaining sum over the reduced momentum reads:
\be
f_\bn(\e) =  \frac 1 N \sum_{\bq} {\rm e}^{i\bq\cdot\bn}\,{\rm Tr}\,\hat G_{\bq}^{(0)}(\e),
\ee
what can be routinely approximated by the following integral (see also Eq.~\ref{int1}):
\bea
f_\bn(\e) & \approx & \frac{4\e}{q_{max}^2}\int_0^{q_{max}}\frac{J_0(qn)q dq}{\e^2 - \e_q^2} \nn\\
& \approx & \frac{4\e}{q_{max}^2}\int_0^{\infty}\frac{J_0(qn)q dq}{\e^2 - \e_q^2}\nn\\
& = & -\frac{4\e}{W^2}K_0\left(i\frac n{n_\e}\right).
\lb{fne}
\eea%
Here the length scale is set by $n_\e = q_{max}^{-1}W/\e$, so for $n \gtrsim n_\e$ we have $q_{max}n \gg 1$ and it is justified
to extend the integration limit to infinity, transforming the Bessel function $J_0$ into the Macdonald function $K_0$ we can employ its asymptotics: $K_0(ix) \approx \sqrt{i\pi/(2x)}\,{\rm e}^{-ix}$ for $x \gg 1$, see \cite{AbSt}.

Using the above results, and summing over $\bn$, Eq.~\ref{pair}, we can take into account that
$f_\bn$ in $A_{\bn}$ varies very slowly at the lattice scales, $n \sim a \sim q_{max}^{-1}$ (as well as ${\rm e}^{i\bq\cdot\bn}$ at $q \ll q_{max}$), so that averaging of $\s_{\bn}$ follows the rules:
$\langle\s_{\bn}\rangle = 0$, $\langle
\s_{\bn}\s_{-\bn}\rangle = 1$. This makes the contribution of the first term in the numerator negligible compared to
the second one. The resulting expression of $B_\bq$ turns to be already $\bq$-independent, i.e.~$B_\bq \approx B$, where
\be
B \approx -\frac{4\pi n_\e^2 z_\e}{\sqrt 3 a^2}\int_0^\infty\frac{x  dx}{z_\e - i x {\rm e}^{2ix}},
\lb{Bqi}
\ee
with $z_\e = \tfrac{32}3(\pi\e T_h(\e)/W^2)^2$. The numerical estimate for the integral in Eq.~\ref{Bqi} in assumption
of $|z_\e| \lesssim 1$ shows its absolute value to be $\sim 1$, then the corresponding criterion for GE convergence follows
as:
\be
c\frac{4\pi n_\e^2 z_\e}{\sqrt 3 a^2} \lesssim 1.
\ee
This gives an estimate for the GE convergence range:
\be
|\e - \e_{res}| \gtrsim c^{1/2}\frac{\tilde\o^2}W.
\lb{dpair}
\ee
This is deep within the mobility gap estimated in the T-matrix approximation, Eq.~\ref{Dtd}, so the higher order GE terms
cannot influence the formerly established results stemming solely from the T-matrix. Also the above assumed condition
of $|z_\e| \lesssim 1$ is well confirmed in the range by Eq.~\ref{dpair}.

\section{DOS calculation for hollow position impurities}
\lb{B}

In the presence of h-type AM impurities, the DOS, more precisely the part dominated by the host bands, is conventionally
obtained from Eq.~\ref{rho}. For that one would need the perturbed GF, Eqs.~\ref{se} and \ref{ge}, that can be in the lowest
order in $c$ obtained with the help of the T-matrix, for its the explicit form see Eq.~\ref{tmh1}.
Taking all that on gets for the trace of the locator of the perturbed GF, ${\rm Tr\,}\hat G_{loc}(\e) \equiv g(\e)$,
the following expression
\be
\begin{aligned}
g(\e) & =\frac{2}{\pi q_{max}^2}\,\int_0^{2\pi}d\theta \int_0^{q_{max}}qdq \\
& \times  \frac{\e D - \tilde c\e_q^2}{\left(\e^2 - \e_q^2\right)D - \tilde c\e_q^2\left(\e - \e_q \cos \theta\right)},
\end{aligned}
\lb{b1}
\ee
where $D = D(\e)$ is given by Eq.~\ref{dh}. The integral over the azimuthal variable $3\theta_\bq$ was substituted by $\theta$
and when taking into account also the shift of the upper limit it gives what is stated above.
The angular integration over $\theta$ can be carried out with the help of the standard formula:
\[\int_0^{2\pi} \frac{d\theta}{a - b\cos{\theta}} = \frac{2\pi}{\sqrt{a^2 - b^2}}.\]
The radial integration over $q$ can be processed in terms of the new variable $x = \e_q^2$:
\be
g(\e) = \frac{2}{W^2}\,\int_0^{W^2}\frac{\left(\e D - \tilde c x\right)dx}{\sqrt{(\e^2 - x)(x - x_1)(x - x_2)}},
\lb{b2}
\ee
where the energy dependent roots in the denominator count:
\[ x_{1,2} = \frac{D + 2\tilde c\e \pm \sqrt{D(D + 4\tilde c\e)}}{2\tilde c^2}.\]
The above integral, Eq.~\ref{b2}, can be calculated analytically, after passing from $x$ to the trigonometric variable $u$:
\[ u = \arcsin{\frac{2x - x_1 - x_2}{x_1 - x_2}},\]
and results in:
\bea
g(\e) & = & \frac{\sqrt{x_1 - x_2}}{W^2}\,\int_{u_1}^{u_2}\frac{(\sin u + \a_1)du}{\sqrt{\sin u + \a_2}} \nn\\
& = & \frac{2\sqrt{x_1 - x_2}}{W^2}\left\{\frac{\a_2 - \a_1}{\sqrt{1 + \a_2}}\left[F\left(\frac{u_2}2|\frac2{1 +
\a_2}\right)\right.\right.\nn\\
& - & \left.F\left(\frac{u_1}2|\frac2{1 + \a_2}\right)\right] - \sqrt{1 + \a_2}\left[E\left(\frac{u_2}2|\frac2{1 + \a_2}
\right)\right.\nn\\
&&\qquad\qquad\qquad \left.\left. - E\left(\frac{u_2}2|\frac2{1 + \a_2}\right)\right]\right\}.
\lb{b3}
\eea
Here $F(x|y)$ and $E(x|y)$ are, respectively, the elliptic integrals of the 1st and 2nd kind \cite{AbSt} and their arguments
include the energy dependent terms:
\begin{align}
\a_1 & = \sqrt{\frac {2D}{D + 4\tilde c\e}}, &  u_1 &=  \arccos \frac{2W^2  - x_1 - x_2}{x_1 - x_2},\nn\\
\a_2 &= \frac{x_1 + x_2 - 2\e^2}{x_1 - x_2}, &  u_2 &= \arccos\frac{x_1 + x_2}{x_2 - x_1}.\nn
\end{align}
The result of Eq.~\ref{b3} permits analytic approximations for the host part of DOS,
$$
\r_h(\e) = \frac 1\pi{\rm Im}\,g(\e),
$$
and, then, similarly for the impurity part of DOS:
$$
\r_{imp}(\e) = \frac 1\pi {\rm Im}\left[\,\frac c{\e - \e_0 - \left(\o\e/t\right)^2 g(\e)}\right].
$$
The resulting total DOS, $\r_{tot}(\e) = \r_h(\e) + \r_{imp}(\e)$, is presented in Fig.~\ref{fig9}. It clearly
displays the contributions from the spectrum branches $E_{^1_2,\bq}$ with van Hove singularities at their special energies
$\e_{res}$ and $\tilde\e_{res}$ and practically restores the unperturbed $\r_0(\e)$ when going with energy beyond the
impurity bands that own widths $W_{imp}$ and $W'_{imp}$.

\end{document}